# Spin-dependent electronic transport in NiMnSb/MoS$_2$(001)/NiMnSb magnetic tunnel junction

Aloka Ranjan Sahoo[1,2] and Sharat Chandra[1,2]

[1]Materials Science Group, Indira Gandhi Centre for Atomic Research, Kalpakkam, 603102, Tamil Nadu, India

[2]Homi Bhabha National Institute, Mumbai, 400094, India

Corresponding author's email: alokranjanctc@gmail.com, sharat@igcar.gov.in

Short title:

*Abstract*

Half-metallic Heusler alloy compounds with curie temperatures above room temperature are suitable candidate electrode materials for injecting large spin-polarised charge carriers into the semiconducting barriers at the ferromagnet semiconductor junction to obtain highly spin-polarised current. Combining the density functional theory and non-equilibrium Green's function method, the electronic structure, spin-dependent electron transport in NiMnSb/MoS$_2$(001)/NiMnSb is studied. The possibilities of injecting 100% spin polarised electron into MoS$_2$ using half metallic NiMnSb as an electrode, the layer dependent, and the effect of the type of interface on electronic structure and spin-transport properties in magnetic tunnel junction devices are studied. We show that the half-metallicity of NiMnSb(111) is preserved at the interface between the half-Heusler alloy NiMnSb and MoS$_2$. NiMnSb keeps a fully spin-polarised state in the majority spin channel at the interface between NiMnSb and MoS$_2$, injecting fully spin-polarised electrons into the semiconductor. The device based on NiMnSb/MoS$_2$(single layer)/NiMnSb has a metallic interface. Metal-induced states in the spin-majority channel of MoS$_2$ are seen after making an interface with half-metallic NiMnSb. In contrast, the NiMnSb/MoS$_2$(three layers)/NiMnSb interface with a multilayer of MoS$_2$ has a bandgap region, and electrons can tunnel through the junction. The Mn-S interface is more conducting than the Sb-S interface due to the strong bonding of Mn and S atoms at the Mn-S interface.

## INTRODUCTION

Following the discovery of giant magneto-resistance and the observation of spin injection and detection in metallic multilayer [1], and thereafter electronic transport in layered materials [2,3] a field of research known as "spintronics" emerged dedicated to creating practical devices that exploit "spin" degrees of freedom of the electron for electron transport in semiconductor-based electronic devices [4]. Using the electronic spin and charge leads to increased data processing speed and low power consumption for spintronic devices compared to conventional electronics that only use electronic charge. Significant enhancement is possible in spintronic devices over traditional electronic devices, where electron spin plays a vital role in the transfer and storage of information [5].

Semiconductor-based spintronics systems have garnered particular research interest because semiconductors can be integrated within modern-day electronics, thus improving device efficiency and storage capacity. The fundamental goal in spintronics is to control the electron spin and electrically tune highly efficient spin injectors and detectors, preferably compatible with nanoscale electronics for valuable applications of spintronics devices. Several experiments and theoretical investigations are carried out to find novel electrode material and two-dimensional (2D) spacers for new spin-valve (SV) devices for next-generation spintronics applications [6,7] . The primary focus remains on the injection and transportation of spin-polarised carriers in heterointerface nanoscale devices, the optimisation of electron spin lifetime, the manipulation and detection of electron spin, and the detection of spin coherence in nanoscale devices.

The magnetic tunnel junction (MTJ) spin-valve device offers a way to understand the mechanism behind the transport of spin-polarised electrons through the insulating or semiconducting barrier sandwiched between two ferromagnetic electrodes. When voltage is applied across the electrodes in MTJ, electrons from the left electrode tunnel through the insulating/semiconducting barrier to reach the right electrode. Depending on the relative magnetisation of the electrodes, the device configuration allows electrons with one spin to propagate through the device (offers a low resistance path), and the opposite spin is reflected or scattered away (offers a high resistance path). This occurs because spin propagation depends on the alignment of the magnetic moments in the ferromagnet. Thus, a "spin-polarised current" is produced. However, generating large spin-polarised currents requires ferromagnetic material having a large spin-polarised state at the Fermi level. Spin valves with normal ferromagnetic material are inefficient or require large polarising magnetic fields.

In the limit of 100% spin-polarization when the spin current is only due to one spin direction, The MTJ allows only one particular spin to propagate and does not allow or restrict propagation for an opposite spin. In this scenario, the MTJ device behaves as an electron spin switch. Thus, finding some magnetic materials with a large polarisation at the Fermi level is preferred. Materials having such a fully polarised state at the Fermi level were theoretically predicted for a new class of materials called half-metallic ferromagnets (HMF) [8–10]. These materials possess unique DOS characteristics; one of the spin channels is metallic, and the other spin channel is semiconducting or insulator. A few theoretically found first half-metallic materials are the LaSrMnO$_3$ perovskite alloy [11], the NiMnSb half-Heusler alloy [8,12], Fe$_3$O$_4$ [13], and CrO$_2$ [14]. The LaSrMnO$_3$ has a low Curie temperature (around 300 K). On the contrary, the Curie temperatures of NiMnSb, Fe$_3$O$_4$, or CrO$_2$ alloys are much larger than at room temperature. Several other HMFs were proposed recently [15–19]. Due to their total spin polarisation, these materials can improve the performance of magneto-resistive devices such as spin valves and magnetic tunnel junctions (MTJ). NiMnSb is an HMF with a metallic majority spin channel and a semiconducting minority spin channel. The high spin polarisation of DOS at the fermi level and Curie temperature (T$_C$) above room temperature at 730 K makes it a suitable candidate electrode material for injecting spin-polarised current into semiconductors in MTJ [20–22].

Oxide-based wide band gap MgO and Al$_2$O$_3$ serving as a 2D spacer in MTJ has been proposed; it does not give desired high magnetoresistance (MR) in vertical Spin-valves [23–28]. Atomically thin Two-dimensional (2D) layered materials offer a unique potential for spintronics devices due to remarkable properties such as long spin-coherence lengths [29,30], spin-polarised tunnelling [24], Inversion of spin signal and spin filtering in ferromagnet Hexagonal Boron Nitride(h-BN)-graphene van der Waals heterostructures. [31] strong spin−orbit coupling (SOC) [32], and spin-momentum locking [33].

Graphene and h-BN-based 2D material as a spacer have been studied [34–36] as low resistance barriers. However, the difficulty in opening a band gap in gapless graphene and modulating the band gap in h-BN is difficult. In contrast, using 2D MoS$_2$, whose band gap can be tuned relatively easily [37,38], is an alternative candidate to use as a spacer in magnetic tunnel junction devices [39–41]. The transition metal dichalcogenide (TMDs) $MoS_2$ has a layered structure with Mo sandwiched between two hexagonal planes of S in a trigonal prismatic arrangement. The intra-layer bonding of S-Mo-S is covalent, and inter-layer S-Mo-S sandwiches interact via weak van der Walls force [42,43].

$MoS_2$, in its 2D monolayer form, offers a high current on-off ratio ($\approx 10^8$), high carrier mobility (200 $cm^2$/Vs), and high thermal stability in the field effect transistor [44]. The indirect band gap of 1.2 eV in bulk $MoS_2$ changes to the direct band gap of 1.9 eV in monolayer $MoS_2$ [37,45–48], High spin-orbit coupling (SOC) and breaking of inversion symmetry in monolayer $MoS_2$ open valley degrees of freedom and corresponding field valley electronics [49].

NiMnSb is a ferromagnetic half-metal with a metallic majority spin channel and a semiconducting minority spin channel. The high spin polarisation of DOS at the fermi level and Curie temperature ($T_c$) 730 K above room temperature makes it a suitable candidate electrode material for injecting spin-polarised current into semiconductors in MTJ [21,22,50]. The crystal structure of bulk NiMnSb is face-centered cubic ($F\bar{4}3m$) consisting of four sub-lattices Ni at (0,0,0), Mn at (0.25,0.25,0.25) and Sb at (0.75,0.75,0.75) and vacancies at (0.5, 0.5, 0.5). The calculated lattice parameters a = b = c = 5.89 Å. The bulk crystal has a layered structure, where Ni atoms tetrahedrally surround Mn and Sb atoms.

Efficient spin injection and giant magnetoresistance using $MoS_2$ as a spacer and Fe(001) as a ferromagnetic electrode have been reported by K.Dolui et al. [39] For junctions of greater thickness, they found a maximum MR of ∼300%, which remains robust with the applied bias as long as transport is in the tunnelling limit. Khaldoun et al. have reported considerable MR in planar Fe/MoS$_2$ junction using MoS$_2$ nanoribbons [51] where the electron transport is along the lateral direction of MoS$_2$. Experimentally, Han-Chun demonstrated significant transverse MR using Fe$_3$O$_4$ as an electrode and MoS$_2$ as a spacer. Fe$_3$O$_4$ keeps a nearly fully spin-polarised electron band at the interface between MoS$_2$ and Fe$_3$O$_4$. Also attempted to fabricate the Fe$_3$O$_4$/MoS$_2$/Fe$_3$O$_4$ MTJs. They observed a clear tunnelling magnetoresistance (TMR) signal below 200 K [40]. Recently, spin injection and MR in MoS$_2$-based tunnel junctions using Fe$_3$Si Heusler alloy electrodes obtained spin injection efficiency (SIE) of about 80% and a maximum MR ratio of ~300% for spacers comprising between three and five MoS$_2$ monolayers. Both the SIE and the MR remain robust at finite bias, namely MR > 100% and SIE > 50% at 0.7 V [52]. NiFe/MoS$_2$ interface has also been investigated; the spin-valve effect is observed up to 240 K, with the highest magnetoresistance (MR) up to 0.73% at low temperatures [41], and also from the first-principle electron transport calculations, which reveal an MR of ∼9% for an ideal Py/MoS$_2$/Py junction.

The interface between the ferromagnetic electrode material and the spacer plays a crucial role in electron transport across the interface and, hence, in the functionality of the nanoscale device. Several Theoretical investigations and theoretical studies have reported that the Half-metallicity of Heusler alloy is lost at the interface between Heusler alloy and the spacer due to symmetry breaking at the interface depending on the surface orientation and the type of bonding at the interface [10,53–57]. Thus, the electronic and magnetic properties of the interface are essential for studying spin-dependent electron transport.

In this work, we have done a theoretical study of electron transport in MTJ formed by half-metallic NiMnSb as an electrode and 2D MoS$_2$ as a spacer. We have discussed the electron transport along the (0001) direction of MoS$_2$ with different thicknesses (Single layer and three-layer MoS$_2$) and with two different types (Mn-S and Sb-S atoms making bonds at the interface) of interface geometries. We have shown that the device based on NiMnSb/MoS$_2$(SL)/NiMnSb has a metallic interface. In contrast, a device with a multilayer of MoS$_2$ NiMnSb/MoS$_2$(3L)/NiMnSb, the half-metallic NiMnSb keeps a fully spin-polarised state in the majority spin channel at the interface between NiMnSb and MoS$_2$, injecting fully spin-polarised electrons into the semiconductor.

## Computational details

Density functional theory (DFT) and non-equilibrium Green's function formalism (NEGF) are used for electronic structure calculations and spin transport calculations, respectively, as implemented in the QuantumATK Atomistic Simulations Software package [58–60] . The electronic structure calculations for the bulk MoS$_2$, NiMnSb, and NiMnSb/MoS$_2$ interface are carried out using DFT and the spin transport across the NiMnSb(111)/MoS$_2$(0001)/NiMnSb(111) junctions are calculated with NEGF formalism. The Linear combination of atomic orbitals (LCAO) method for the basis set and FHI double-zeta polarised norm-conserving pseudo-potentials is used for all the elements with a density mesh cut-off of 150 Hartree [61,62] . The generalised gradient approximation (GGA) function of the Perdew-Burke-Ernzerhof (PBE) form is used to describe the exchange-correlation effects [63]. Energy tolerance of 10$^{-6}$ eV is set for achieving the self-consistency of electronic energy minimisation, and the force tolerance for ionic relaxations is fixed at 10$^{-2}$ eV/Å for all the atoms. For the electronic structure calculation of bulk NiMnSb, the Brillouin zone is sampled with a K-mesh grid of 13×13×13 using a Gamma-centered Monk-horst pack scheme [64]. The two-probe models used in our calculations are shown in Figure 1. The heterostructure junction is made by sandwiching a $MoS_2$ slab of (0001) crystallographic orientation between two NiMnSb (111) electrodes [65]. The NiMnSb/$MoS_2$/NiMnSb interface geometries are made by applying strain on $MoS_2$ to match the in-plane lattice parameter of NiMnSb, resulting in a lattice mismatch of 0.3%. A 7×7×1 K-mesh grid with a Gamma-centered Monk-host pack scheme is used to sample the 2D Brillouin zones of the interface geometry and the MTJ. The heterostructure geometries of NiMnSb(111)/$MoS_2$(0001)/NiMnSb(111) junction with the $MoS_2$ as spacer having thickness of 1 layer (1L) and 3 layers (3L) of $MoS_2$ are modelled. The in-plane lattice parameters of the heterostructure are a=b = 8.34 Å and α = 60º. The study considers two different heterostructure interface geometries based on the type of atom present at the interface of NiMnSb and $MoS_2$. The first heterostructure geometry has Sb-S atoms forming contact with $MoS_2$ on both interface sides (heterostructure-1). The second heterostructure has Sb-S atoms forming a contact on one side and Mn-S atoms forming the contact on the other side of $MoS_2$, forming the interface (heterostructure-2). All the geometries are relaxed until the forces on each atom are less than the stopping criterion of $10^{-2}$ eV/Å. The electrode needs to have sufficient electrode layers to screen any perturbation arising due to any scattering at the interface of NiMnSb and $MoS_2$ from reaching the electrodes. Thereby ensuring that the electrode length is sufficient to replicate the bulk behaviour of NiMnSb. Grimme-D2 correction is included in the calculation to account for the van der Walls correction required to take care of the interaction between the $MoS_2$ layers [66]. The geometry of the MTJ is shown in Figure 1 for Sb-S atoms on both sides of the interface (heterostructure-1) Figure 1 (a), Sb-S atom on the left side, and Mn-S on the right side Figure 1 (b) (heterostructure-2).

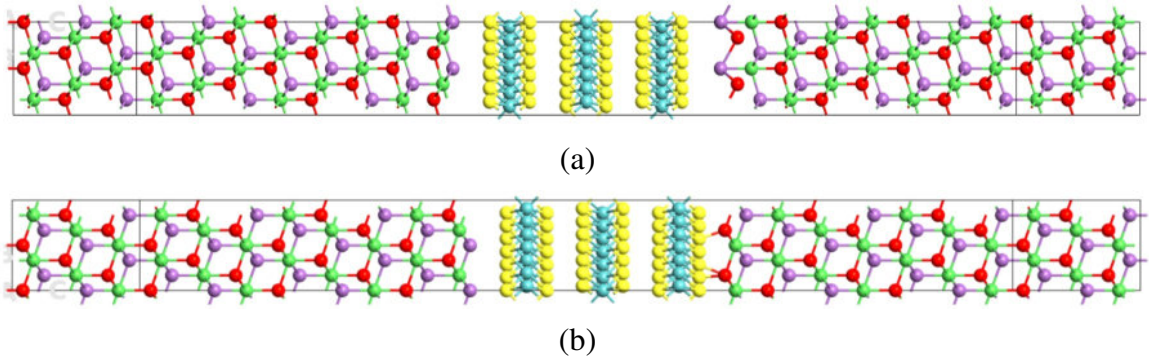

(a)

(b)

*Figure 1 Heterostructure geometry with Sb-S at both sides of the interface (a) and Sb-S at the left side and Mn-S at the right side at the interface (b) balls of different colours denote atoms red-balls for Manganese, green-balls for Nickel, purple-balls for Antimony, yellow-balls for Sulfur, and Cyan for Molybdenum atoms.*

## Results and Discussions

## Bulk NiMnSb

The crystal structure of bulk NiMnSb is face-centred cubic ($F\bar{4}3m$) consisting of four sub-lattices Ni at (0,0,0), Mn at (0.25,0.25,0.25), Sb at (0.75,0.75,0.75) and vacancies at (0.5, 0.5, 0.5). The calculated lattice parameters a = b = c = 5.89 Å. The bulk crystal has a layered structure, where Ni atoms tetrahedrally surround Mn and Sb atoms. The electronic density of states (DOS) of bulk NiMnSb is shown in Figure 2(a), and the atom-projected density of states (PDOS) is shown in Figure 2(b). Figure 2(a) shows the half-metallic nature of NiMnSb, which has a semiconducting band gap of 0.8 eV in the spin-down channel, and a metallic nature is seen along the spin-up channel. From the Mulliken population, the magnetic moment of Ni, Mn, and Sb are 0.19 $\mu_B$, 4.06 $\mu_B$, and -0.25 $\mu_B$, respectively, bringing the total magnetic moment per unit cell to 4.0 $\mu_B$ as per the signature characteristic of half-metallic Husler alloy. In the bulk unit cell, the spins of Ni and Mn are ferromagnetically aligned, while the spins of Sb are aligned in antiferromagnetic order. For studying the transport properties in the heterostructures, the NiMnSb surface is cleaved along the (111) direction and the MoS$_2$ surface is cleaved along the (0001) directions for forming the NiMnSb/MoS$_2$/NiMnSb junction, both the NiMnSb(111) surface and MoS$_2$ has hexagonal symmetry. The NiMnSb(111) surface has in-plane lattice parameter a = b = 4.17 Å (α = β = 90º, γ = 120º) and MoS$_2$ has in-plane lattice parameter of a = b = 3.16 Å (α = β = 90º, γ = 120º).

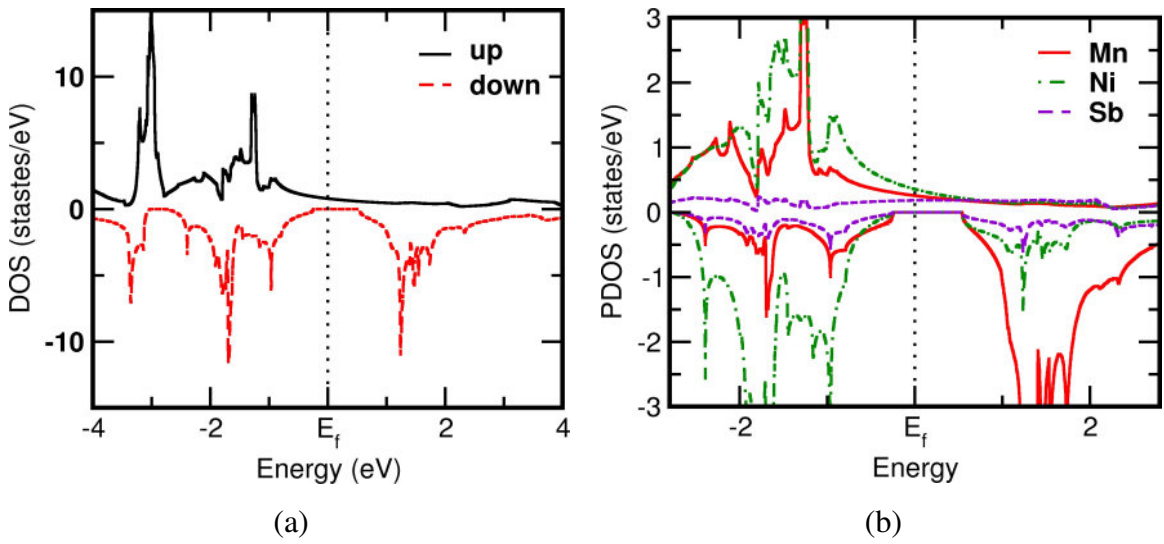

*Figure 2    DOS of bulk NiMnSb (a), PDOS of the NiMnSb electrode (b)*

## Bulk MoS$_2$

The crystal structure of MoS$_2$ has hexagonal symmetry with space group P6$_3$mmc. The MoS$_2$ has a layered structure where the transition metal Mo is sandwiched between two trigonal planes of the chalcogen S atom in a trigonal prismatic arrangement. The electronic structure calculations for MoS$_2$ are done using GGA-FHI norm-conserving pseudopotentials and a double-zeta polarised basis set. The GrimmeD2 corrections to the energy are included for van der Waals corrections in the DFT calculations. The calculated lattice parameters for the MoS$_2$ are a = b = 3.25 Å and c = 12.16 Å. The electronic DOS and bandstructure for MoS$_2$ are shown in Figure 3 (a) and (b), respectively. The calculated electronic band gap for $MoS_2$ is 0.7 eV.

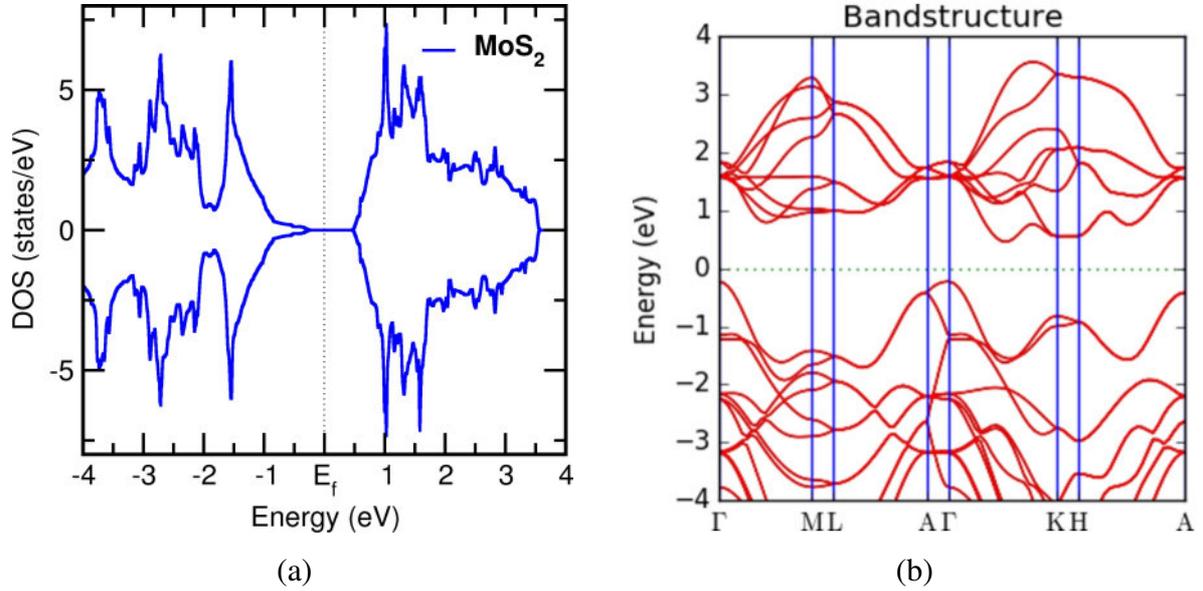

*Figure 3  Electronic structure of bulk $MoS_2$ with GGA-FHI pseudo-potential DOS (a) and Bandstructure (b).*

## Electronic structure and spin transport in Heterostructure-1

*Single layer $MoS_2$ spacer in Heterostructure-1*

First, we discuss the electronic structure of NiMnSb/$MoS_2$(1-layer)/NiMnSb interface in heterostructure-1 before studying the transport properties through it. Figure.4(a) shows the projected density of states of $MoS_2$ in the heterostructure-1. The PDOS of $MoS_2$ at the interface indicates that the semiconducting nature of $MoS_2$ has disappeared, and there are 0.8 states/eV in the spin-up channel at the fermi level. The NiMnSb/$MoS_2$ junction is metallic after forming an interface with NiMnSb. The metallic nature of $MoS_2$ arises from the effect of metal-induced states from the half-metallic NiMnSb due to the hybridisation of Ni-*d*, Mn-*d*, and Sb-*p* orbitals of NiMnSb with Mo-*d* and S-*p* orbitals of $MoS_2$ as shown in Figure.4(b). The contribution to the DOS mainly comes from the Mo-d and S-p orbitals of $MoS_2$ and the Mn-*d*, Ni-*d*, and Sb-*p* orbitals of NiMnSb.

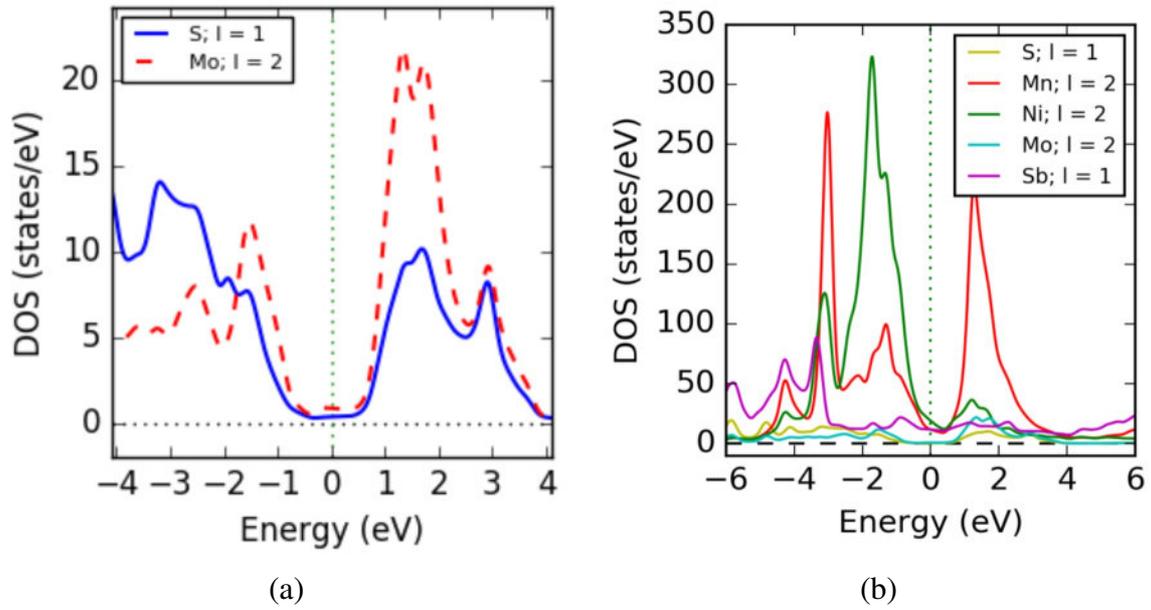

*Figure.4 PDOS for the single layer $MoS_2$ (a) and orbital projected DOS at the interface of $NiMnSb/MoS_2$ in heterostructure-1.*

Figure 5(a) shows the layer-projected density of states (LDOS) for the 3rd and 4th nearest neighbour interface layers (NN) of NiMnSb in the central region of the device. The layer projected DOS of the third and fourth nearest neighbour layers are the same as bulk NiMnSb. The third and fourth nearest layers of NiMnSb from the interface have a semiconducting band gap in the spin-down channel, and the spin channel is conducting in nature. The third and fourth nearest NiMnSb layers have electronic DOS; this shows that the effect of interface layers is restricted to the first two layers of NiMnSb from the interface, forming the central region only. Figure 5(b) shows PDOS for the single-layer MoS2 in the device in anti-parallel configurations. In anti-parallel configurations, both the spin-up and spin-down channels of $MoS_2$ are conducting.

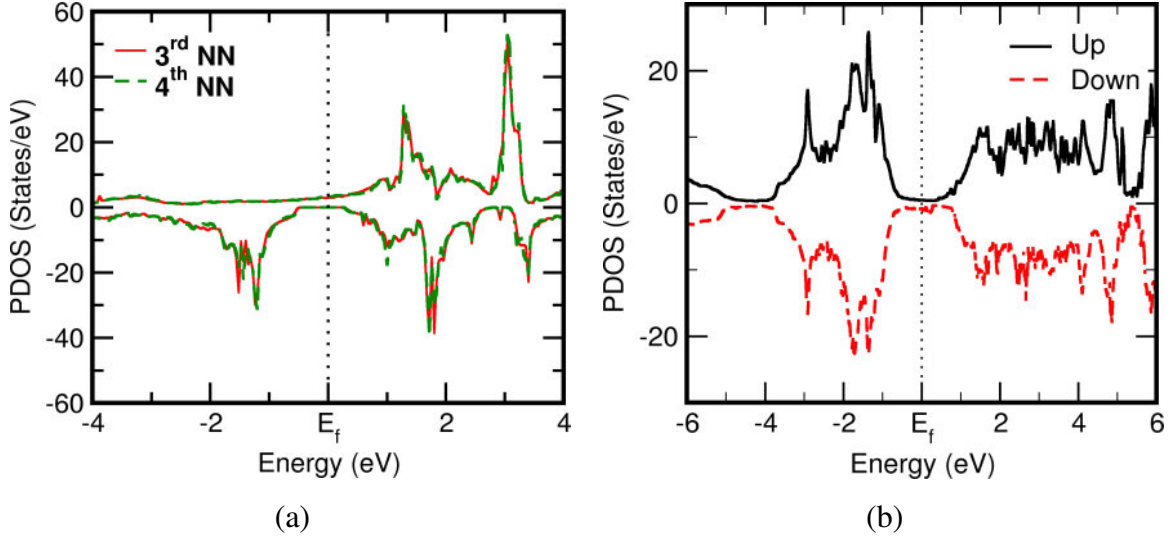

*Figure 5    Layer projected density of states of 3rd and 4th neighbour NiMnSb layers from the interface (a), PDOS of MoS$_2$ in anti-parallel configuration (b) in heterostructure-1.*

The projected local device density of states (PLDOS) of the junction in heterostructure-1 is shown in Figure 6 and Figure 7 For parallel and anti-parallel configurations, respectively. The PLDOS, in which the device DOS is projected over the spatial region of the device along the transport direction, *z*-direction. The transport direction *z* is sliced into different regions, and the variation of the electronic density of states with respect to *z*-direction is shown in the PLDOS. The spatially resolved density of states is shown in the PLDOS using a 2D colour map showing the distribution of electronic states over both energy and real space. Scanning across the z-direction, The PLDOS shows the change in local electronic structure across the electrode, interface, and central scattering region. Brighter red dots show a high number of states, and a dark dot shows a low number of states in the energy range and the atomic layer contributing to the DOS. A dark black region shows the absence of states, indicating a bandgap region for the semiconductor or insulator spacer in the device. The PLDOS for spin-up and spin-down channels and spin-dependent transmission at zero bias in Figure 6(a) and (b) shows that the interface of heterostructure-1 has metal-induced states in MoS$_2$ as seen by the presence of states indicated by blue regions in the local device density of states and finite transmission along the spin-up channel. The spin-down channel has no states available and has zero transmission of electrons. The $MoS_2$ has finite DOS for spin-up electrons and bandgap for spin-down electrons, thus behaves as a half-metallic spacer. At zero bias in anti-parallel configuration, the transmission along both the spin channels is zero, as no states are not available for the spin-up electrons in the left electrode to be transmitted to the right electrode and vice-versa, as shown in Figures 7(a) and (b), respectively.

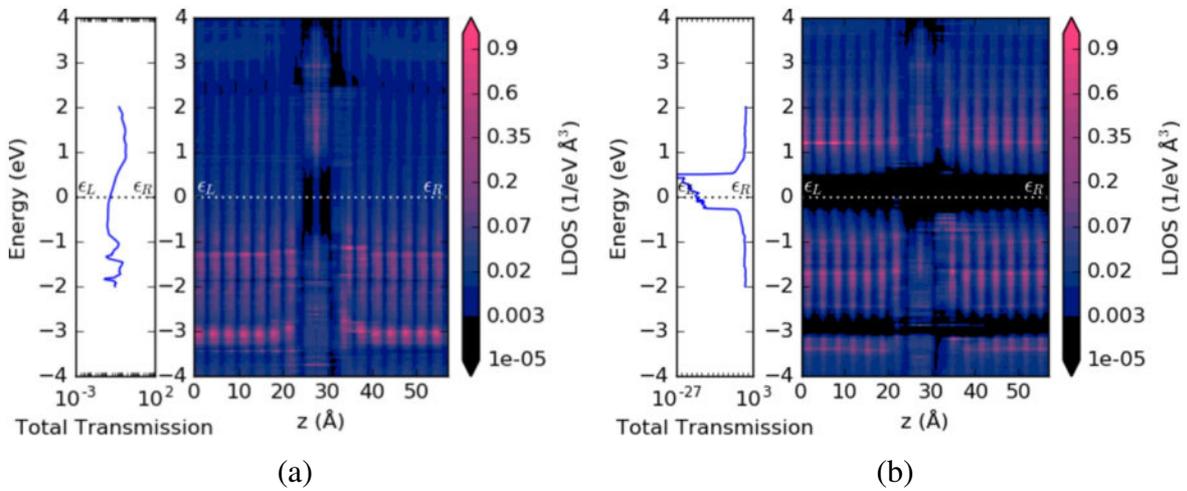

(a)    (b)

*Figure 6    PLDOS of NiMnSb/MoS$_2$(1-layer)/NiMnSb junction (heterostructure-1) in spin-up (a) and spin-down (b) channel for parallel orientation of electrodes.*

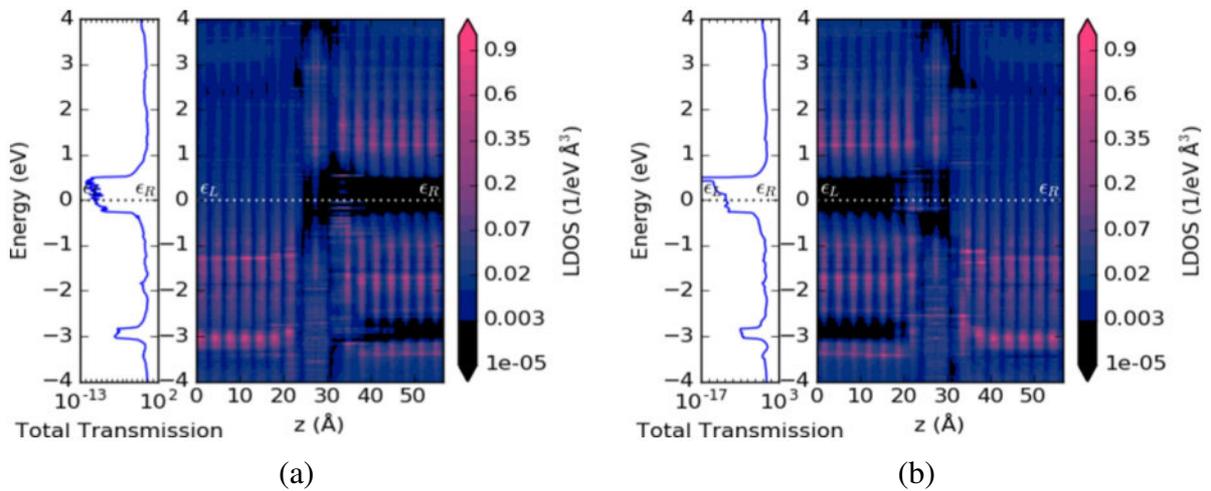

(a)    (b)

Figure 7    PLDOS of NiMnSb/MoS$_2$(1-layer)/NiMnSb junction in spin-up (a) and spin-down (b) channel for anti-parallel orientation of electrodes.

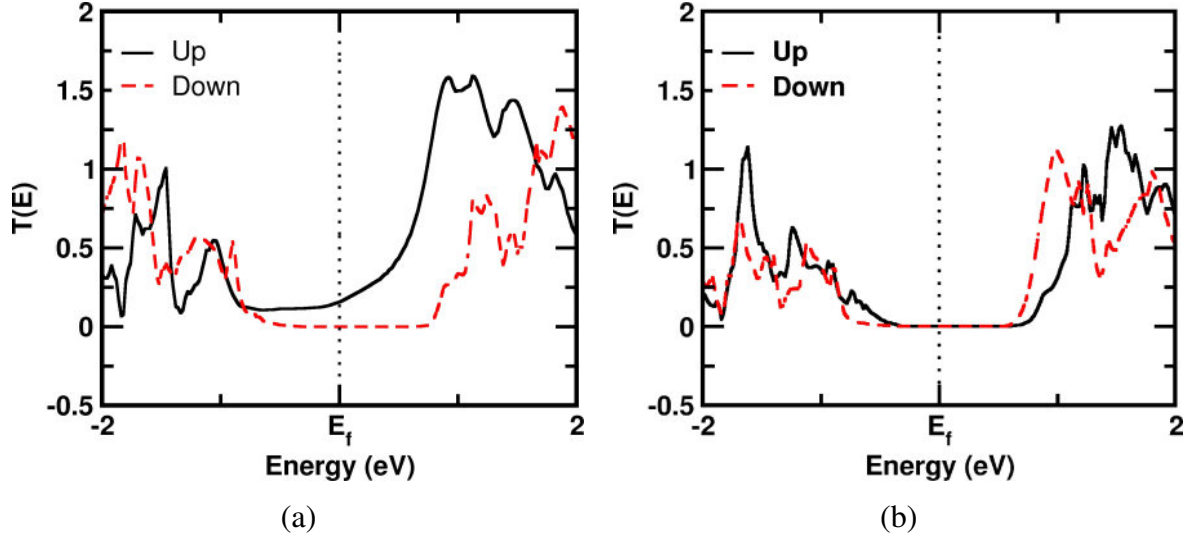

*Figure 8    Transmission spectrum in NiMnSb/MoS$_2$(1-layer)/NiMnSb junction in parallel (a) and anti-parallel (b) orientation of electrodes.*

The spin-dependent electron transmission for parallel and anti-parallel configurations of the device is shown in Figure 8(a) and (b), respectively. For parallel configuration, electron transmission is zero across the spin-down channel, and the spin-up channel shows a finite electron transmission of 0.16. The finite electron transmission in the spin-up channel at zero bias is due to the metallic nature of the spin-up DOS in the junction. For anti-parallel configuration, the transmission at zero bias condition is zero for both the spin channels. The dissimilarity in the transmission spectrum for spin-up and spin-down channels in the anti-parallel configuration is attributed to the asymmetric nature of the junction. The total device density of states (DDOS) for the device in parallel and anti-parallel orientation of the electrodes is shown in Figure 9. The metallic nature of the junction in the spin-up channel and semiconducting behaviour in the spin-down channel are observed in Figure 9(a). For the anti-parallel orientation of the electrodes, the left electrode and right electrode have opposite spin polarisation. Thus, for the total DDOS, the spin-down DOS of the left electrode gets added to the spin-up DOS of the right electrode, as shown in Figure 9(b).

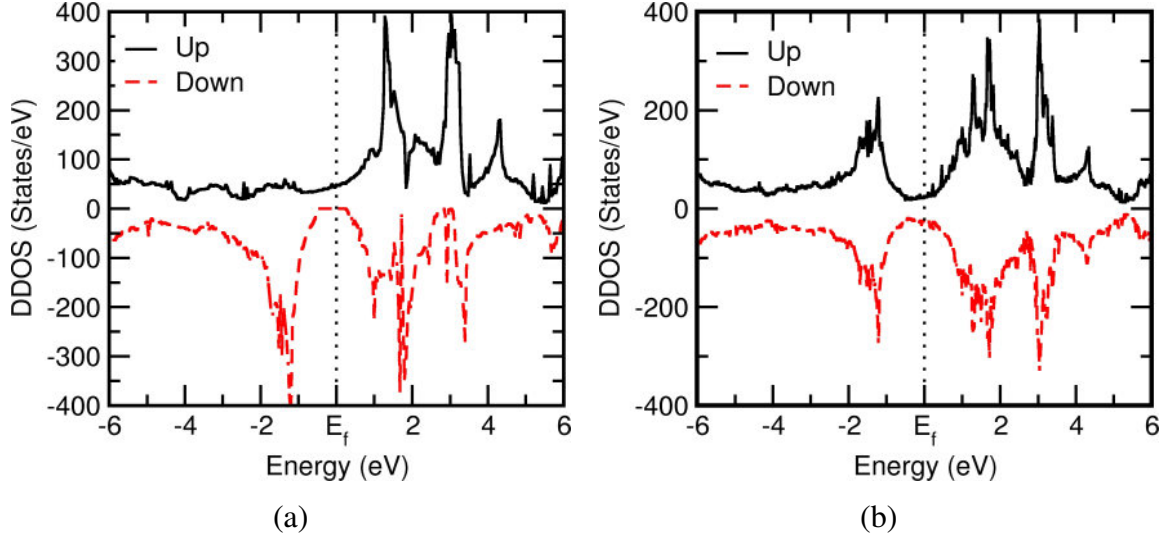

Figure 9   Total device density of states for parallel (a) and anti-parallel (b) orientation of electrodes.

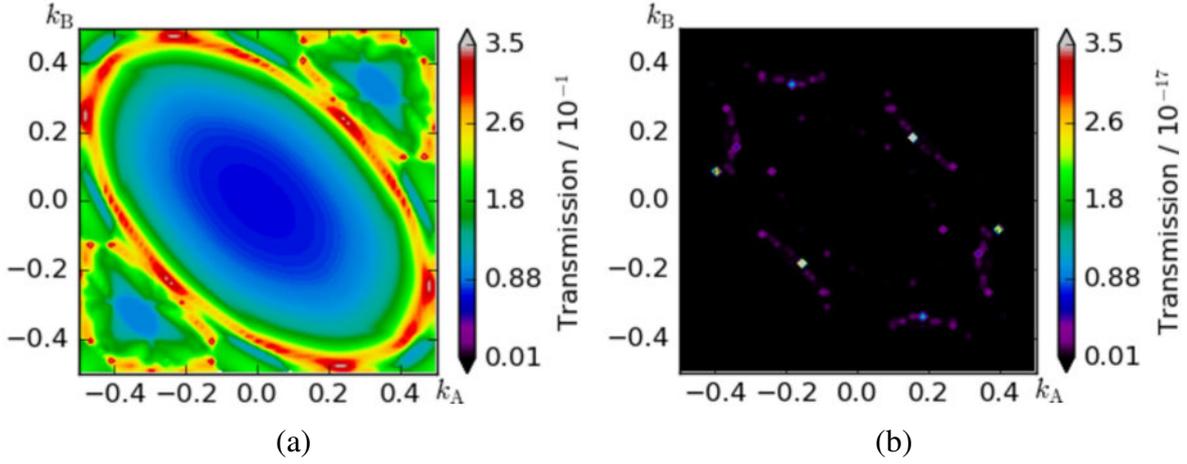

Figure 10   K-resolved transmission spectrum for spin-up and spin-down channels for NiMnSb/MoS$_2$(1-layer)/NiMnSb junction in parallel orientation of electrodes.

The K-resolved transmission spectrum for parallel configuration and anti-parallel configuration of the device under zero-bias conditions is shown in Figure 10 and Figure 11, respectively. The 2D colour map for the K-resolved transmission spectrum shows the contribution to the transmission spectrum from the 2D Brillouin zone at each energy. In the 2D color map, the brighter red dots show a higher transmission coefficient, indicating a higher contribution from the K-point at a particular energy, and a dark dot shows a low transmission coefficient from the K-point at a specific energy, indicating a low contribution from the K-point. The K-resolved transmission spectrum for the spin-up channel in Figure

10(a) shows that the transmission at the Fermi level comes from the entire Brillouin zone. The significant contribution comes from points far away from the Brillouin zone centre. In contrast, the spin-down channel gets no contribution from the Brillouin zone, as shown in dark dots with a transmission coefficient of the order of $10^{-17}$ Figure 10 (b). For the anti-parallel orientation of electrodes, the contribution from the Brillouin zone is zero for both the spin-up and spin-down channels with a transmission coefficient of the order of $10^{-10}$, as shown in Figure 11 (a) and (b). The finite transmission in the spin-up channel, zero transmission for the spin minority channel in the parallel orientation of the electrodes, and zero transmission on both the spin channels in the anti-parallel orientation of the electrodes show an injection of 100% spin polarised electrons in the device.

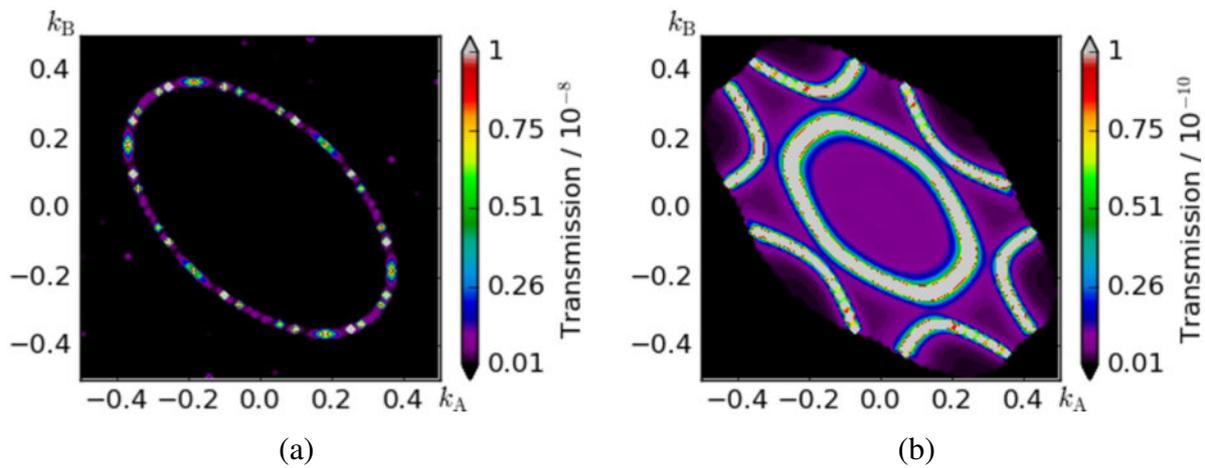

*Figure 11  K-resolved transmission spectrum for spin-up and spin-down channels for NiMnSb/MoS$_2$(1-layer)/NiMnSb junction in anti-parallel orientation of electrodes.*

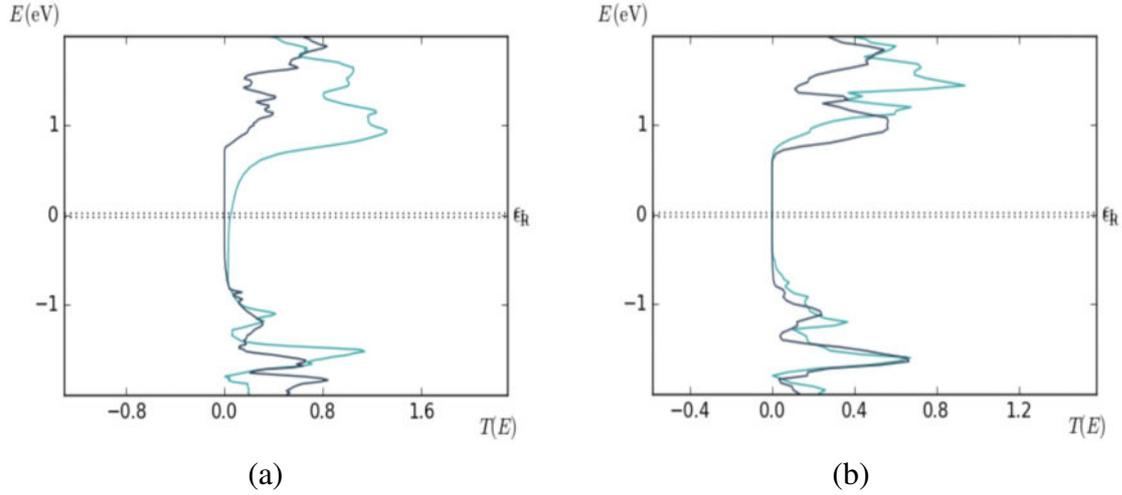

*Figure 12 Transmission spectrum for parallel and anti-parallel configurations for bias 0.05V.*

The transmission spectrum for the device at the bias of 0.05V in parallel and anti-parallel configuration is shown in Figure 12 (a) and (b), respectively; the corresponding K-resolved transmission spectrum for parallel and anti-parallel configuration of the device in the spin-up channel is shown in Figure 13 (a) and Figure 13 (b) respectively. The transmission spectrum for parallel configuration in Figure 12 (a) shows that the chemical potential of the left and right electrodes has shifted and have a potential difference of the applied bias of 0.05V. The spin-up channel has a transmission of 0.05 and only contributes to the transmission spectrum. The transmission of electrons through the junction is only due to the spin-up channel. The K-resolved transmission spectrum for the spin-up channel in parallel configuration is shown in Figure 13 (a). The K-resolved transmission spectrum shows that the contribution to the transmission at finite bias 0.05V comes from the entire Brillouin zone, with the maximum transmission coefficient of order 0.1 coming from the points away from the Brillouin zone centre. For the anti-parallel configuration of the junction, the transmission spectrum in Figure 12(b) and the corresponding K-resolved transmission spectrum in Figure 13 (b) at energy zero show that transmission through the junction in anti-parallel configuration is zero. The finite transmission in spin-up channels only for parallel configuration and zero transmission in both the channels for anti-parallel configuration shows the 100% spin polarisation of the current and the switching nature of the device.

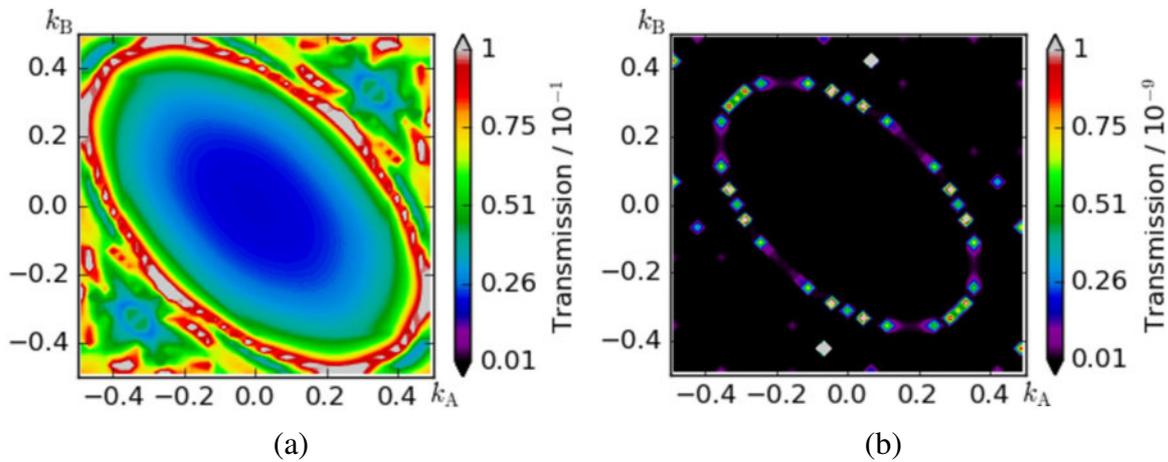

Figure 13    K-resolved transmission spectrum in the spin-up channel for parallel and anti-parallel orientation of electrodes at bias 0.05V.

*Three-layer MoS$_2$ spacer in Heterostructure -1*

For heterostructure-1 with three-layer MoS$_2$ as a spacer in the junction, the total DDOS and layer projected DOS of MoS$_2$ are shown in Figure 15(a) and (b), respectively. The device conducts only in the spin-up channel, and the spin-down channel remains semiconducting. The layer projected DOS of MoS$_2$ in Figure 14 (b) shows that the spin-up DOS for three layers of MoS$_2$ is metallic, whereas the spin-down DOS is semiconducting. The PDOS for the middle layer of MoS$_2$ of single-layer thickness is semiconducting. Thus, the effect of the interface is limited only to the first layer of MoS$_2$ adjacent to the interface on both sides.

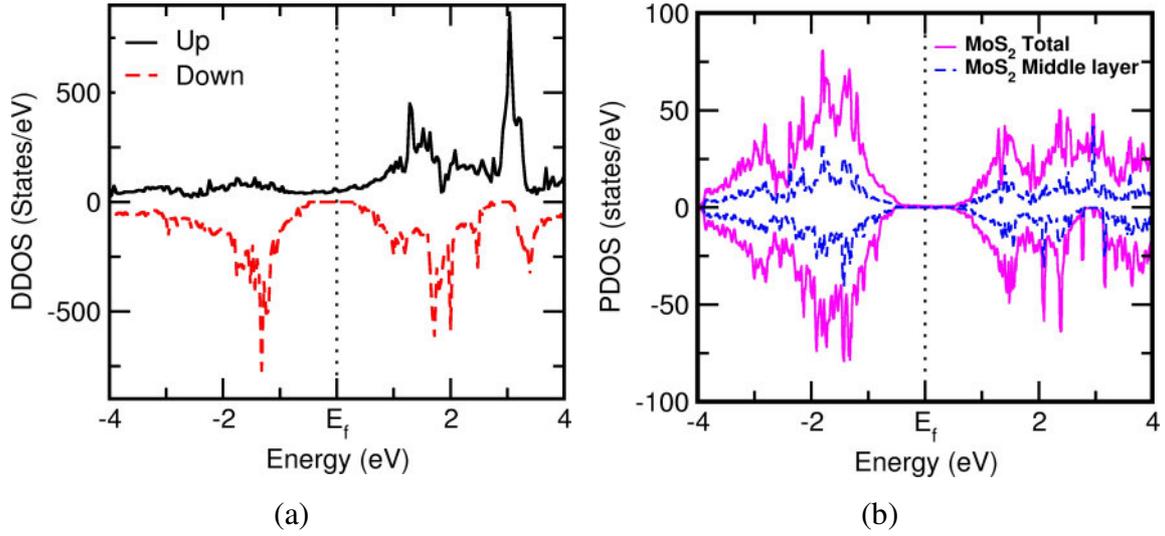

*Figure 14   Total device DOS of NiMnSb/MoS$_2$(3-layer)/NiMnSb junction (a), projected DOS of three-layer MoS$_2$ and middle layer MoS$_2$ (b).*

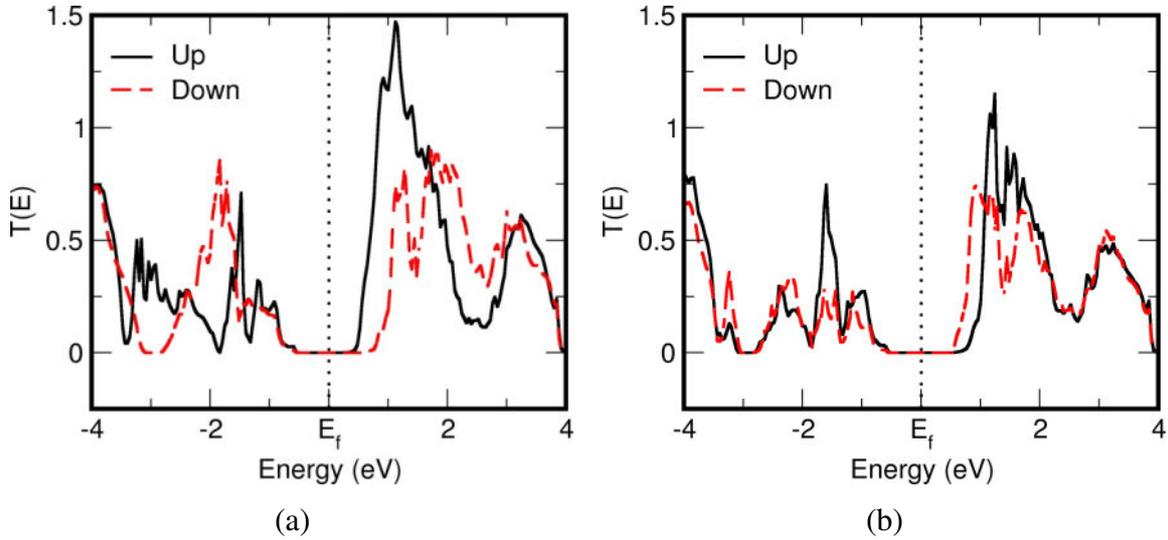

*Figure 15   Transmission spectrum of NiMnSb/MoS$_2$(3-layer)/NiMnSb junction in a parallel configuration(a) and anti-parallel orientation (b).*

The spin-dependent transmission in parallel and anti-parallel configurations at zero bias is shown in Figure 15 (a) and (b), respectively. At zero bias, there is no transmission of electrons for both parallel and anti-parallel configurations, as no states are available for the conduction of electrons. The transmission in the anti-parallel configuration shows zero transmission in both the spin channels, and the transmission spectrum is similar in both the spin-up and spin-down channels. The K-resolved transmission spectra at zero bias for spin-up and spin-down channels in Figure 16 show the contribution to the transmission of electrons from the 2D Brillouin zone. For the spin-up channel, a maximum transmission coefficient of the order of $5\times10^{-4}$ comes from six pockets of the 2D Brillouin zone, and for the spin-down channel, the transmission coefficient is of the order of $10^{-19}$.

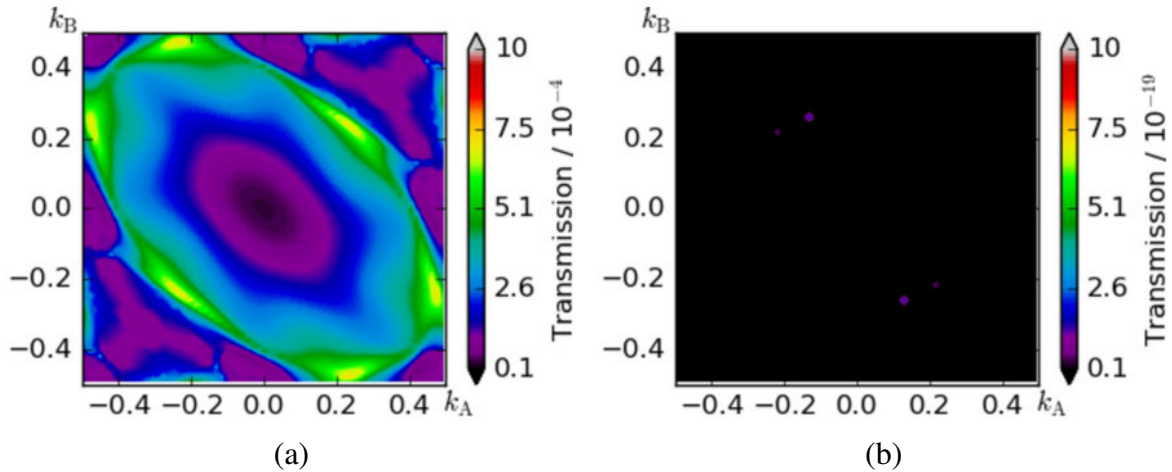

(a) (b)

*Figure 16  K-resolved transmission spectrum in NiMnSb/MoS$_2$(3-layer)/NiMnSb junction for spin-up and spin-down channel at zero bias in a parallel configuration.*

The PLDOS in Figure 17, where the local density of the state is projected on the length of the device along the transport $z$-direction, shows states available from various layers of the device. The PLDOS for parallel configuration shows that at zero bias, the MoS$_2$ layer adjacent to the interface, the spin-up DOS touches the Fermi-level, as seen by the blue dots in the PLDOS showing the low presence of states. In contrast, no states are available for the conduction of electrons in the spin-down channel from the three layers of the MoS$_2$.

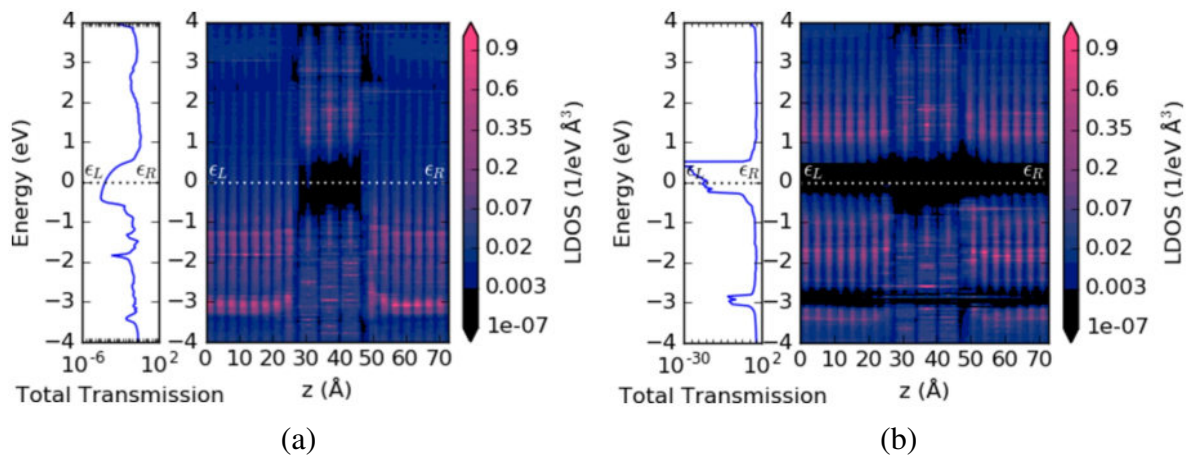

(a) (b)

*Figure 17  PLDOS of NiMnSb/MoS$_2$(3-layer)/NiMnSb junction in the spin-up channel (a), spin-down channel (b),*

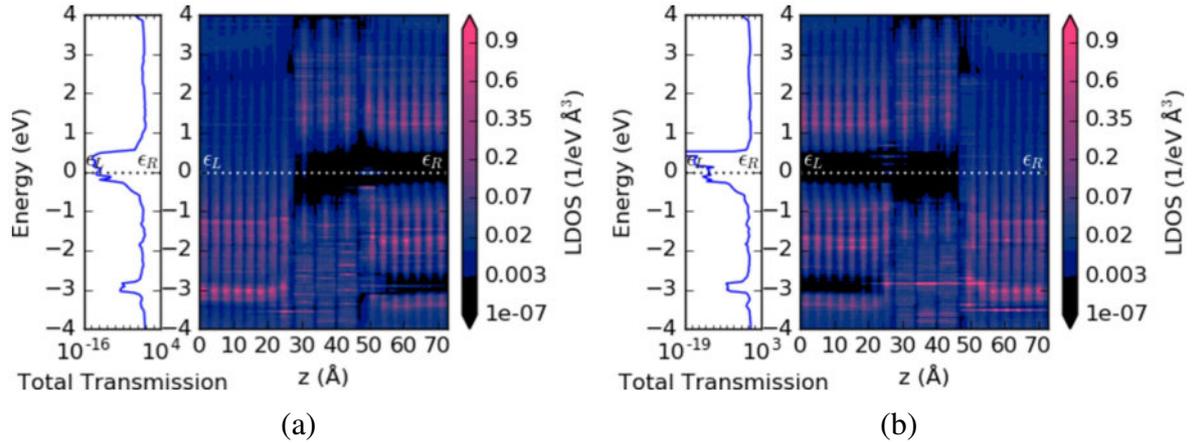

*Figure 18 PLDOS of NiMnSb/MoS$_2$(3-layer)/NiMnSb junction along the spin-up channel (a) and spin-down channel (b) in anti-parallel orientation.*

The PLDOS for the anti-parallel configuration of the junction in Figure 18 Shows that the electron transport from left to right or right to left of the junction is zero due to the unavailability of states in the left/right electrodes. Therefore, the transmission through the junction is zero at zero bias. The K-resolved transmission spectra depict no significant contributions from the 2D Brillouin zone, as shown in Figure 19, with a transmission coefficient of the order of $10^{-10}$.

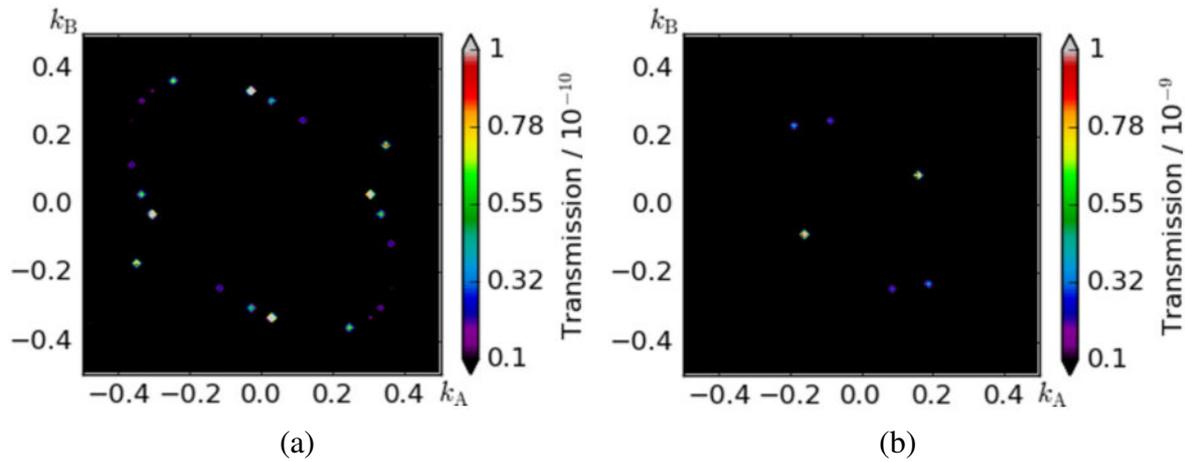

*Figure 19 K-resolved transmission spectrum in spin-up and spin-down channels at zero bias in anti-parallel configuration.*

# Electronic structure and spin transport in Heterostructure 2

*Single layer MoS$_2$ spacer in heterostructure-2*

Figure 20(a) shows the orbital PDOS of the interface and PDOS for single layer of MoS$_2$ in the heterostructure-2. The orbital PDOS shows that the *d*-orbitals of the transition metals Ni, Mn, Mo, and *p*-orbitals of Sb and S contribute to the finite transmission in the spin-up channel. The interface is metallic after forming interface with MoS$_2$, as shown in Figure 20(b)

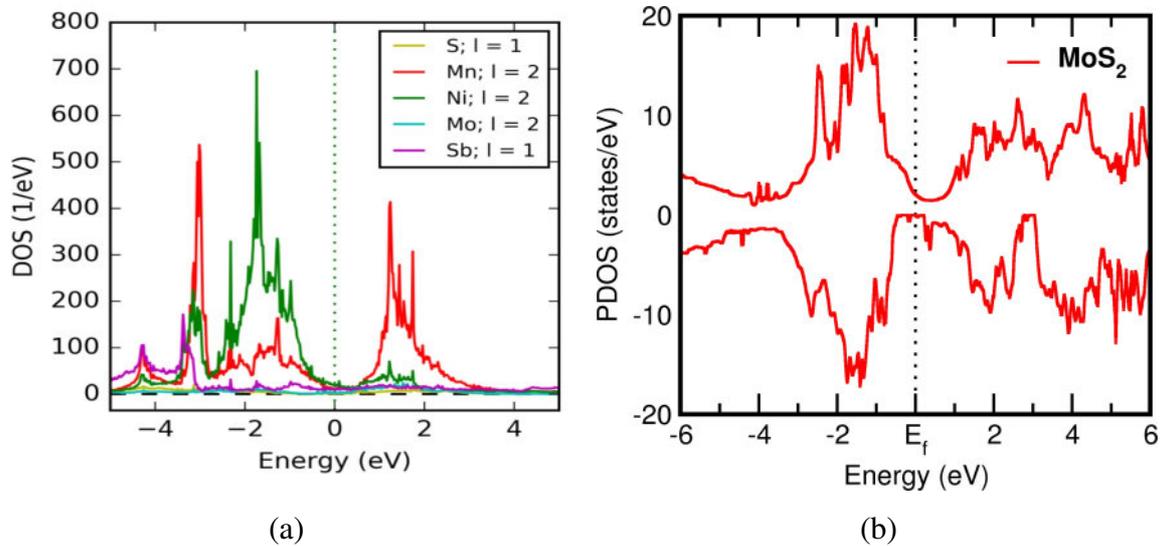

(a)                 (b)

*Figure 20    The orbital projected DOS of the interface of NiMnSb/MoS$_2$/NiMnSb heterostructure-2 and PDOS for MoS$_2$.*

Figure 21 (a) and Figure 21 (b) show the transmission spectrum for the device at zero bias in parallel and anti-parallel orientation, respectively. The transmission spectrum of the device for electrodes oriented parallel to each other shows that there is finite transmission of electrons in the spin-up channel and zero transmission of electrons in the spin-down channel at the fermi level. The transmission spectrum of the device when electrodes are oriented anti-parallel shows electron transmission in spin-up and spin-down channels are similar and have zero transmission at the fermi level for both spin channels.

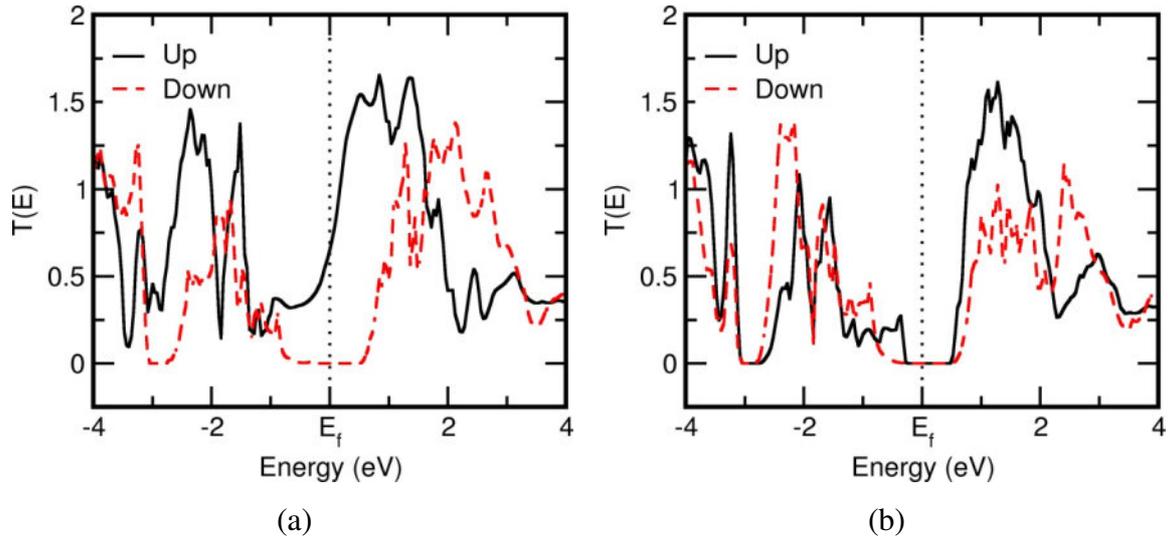

(a)                                     (b)

*Figure 21    Transmission spectrum along spin-up and spin-down channel of NiMnSb/MoS$_2$/NiMnSb (heterostructure-2) in parallel and anti-parallel orientation of electrodes.*

For spin transport across the NiMnSb/MoS$_2$(1-layer)/NiMnSb junction (heterostructure-2), when both NiMnSb electrode layers have the same spin orientation (parallel), Figure 22(a) and Figure 22(b) shows the PLDOS and spin-dependent electron transmission spectrum of the device projected along the length of the device in the transport *z*-direction for both spin-up and spin-down channels. Metal-induced states are present in MoS$_2$ in the spin-up channel. At zero bias, there is a finite (0.65) transmission of electrons in the spin-up channel, as seen in Figure 22(a) and no transmission of electrons in the spin-down channel, as there are no states available, as seen in the black region in the PLDOS in Figure 22(b). For the anti-parallel configuration where the left and right NiMnSb electrodes are oriented opposite to each other, PLDOS and spin-dependent transmission spectrum projected along the length of the device are shown in Figure 23(a) and (b), respectively. Figure 23(a) and (b) show that the transmission of electrons in both spin-up and spin-down channels is zero for the anti-parallel orientation of NiMnSb electrodes, as no states are available for electron transmission.

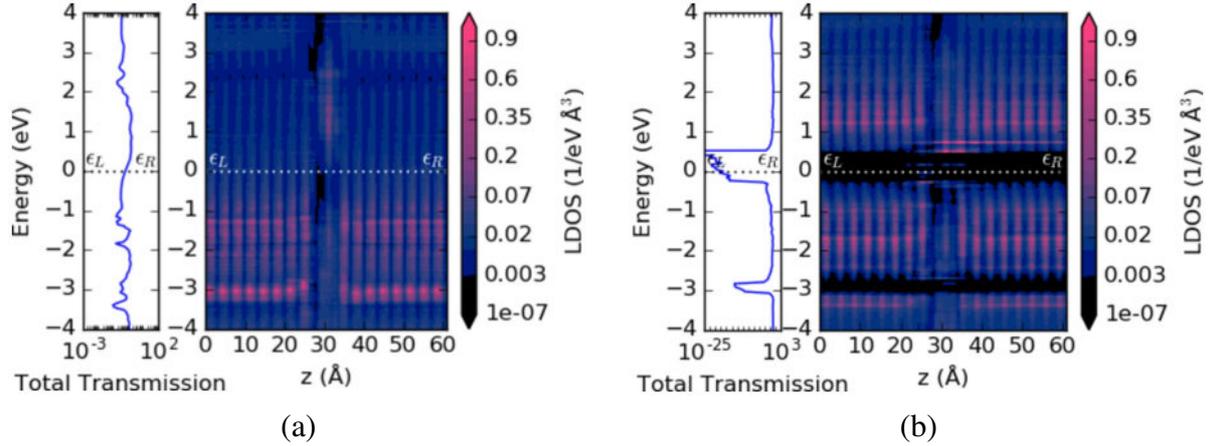

*Figure 22 PLDOS for NiMnSb/MoS$_2$/NiMnSb junction (heterostructure-2) in the parallel orientation of electrodes.*

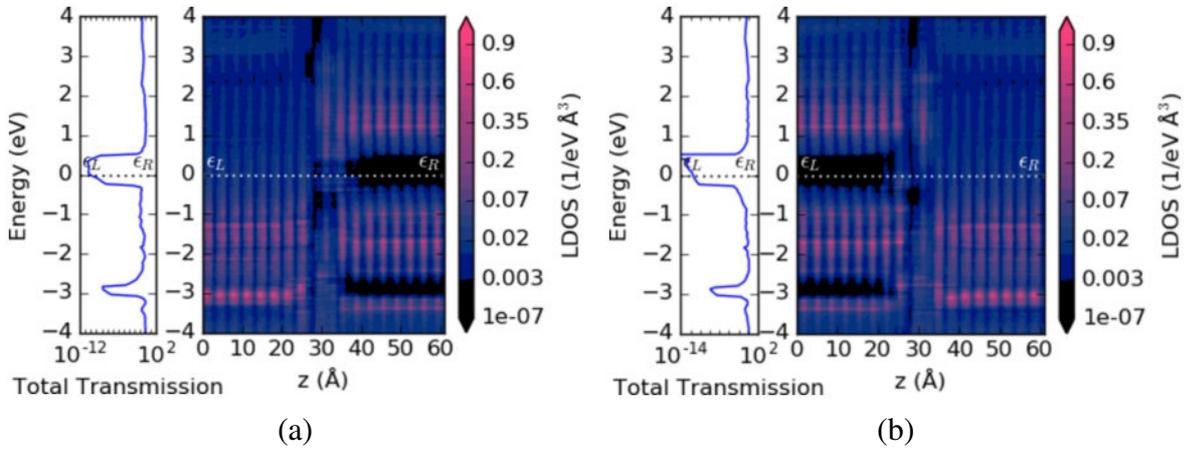

*Figure 23 PLDOS for NiMnSb/MoS$_2$/NiMnSb junction (heterostructure-2) for anti-parallel orientation of electrodes.*

Figure 24 The K-resolved transmission spectrum at zero bias shows the contribution to the transmission spectrum from the entire 2D Brillouin zone at the fermi-level. The k-resolved transmission spectrum shows that the contribution to the transmission spectrum comes from zones away from the Brillouin zone centre, notably six pockets far away from the centre, forming a hexagonal ring structure. There are no contributions to the transmission spectrum from the Brillouin zone for the spin-down channel at the fermi-level in the spin minority channel with a transmission coefficient of the order of $10^{-16}$.

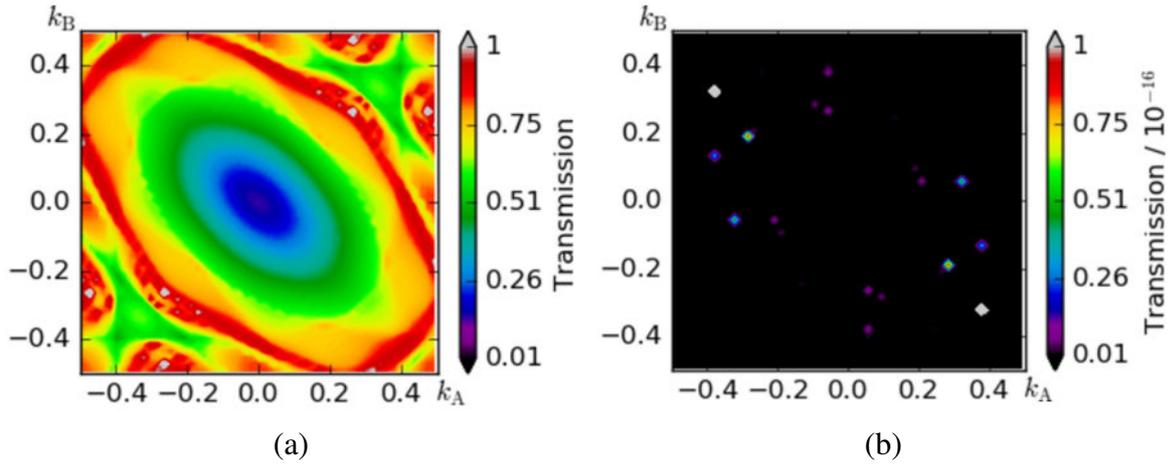

*Figure 24    K-resolved transmission spectrum for up-spin and down-spin channels in NiMnSb/MoS$_2$/NiMnSb junction (heterostructure-2).*

In the anti-parallel configuration of the device, the left electrode has the spin-up electrons as the majority charge carrier and the right electrode has spin-down electrons as the majority charge carrier; therefore, in the resultant device density of states, as shown in Figure 25 (a), the black line shows the contribution from spin-up electrons, which are the majority of charge carriers for the conducting channel of the left electrode, and the red line indicates the contribution from the right electrode, where spin-down electrons are the majority of charge carriers for the conducting channel. The PDOS of MoS$_2$ in Figure 25(b) shows the spin polarisation of spin-up and spin-down electrons of MoS$_2$. The K-resolved transmission spectra are identical for both spin-up and spin-down channels, as shown in Figure 26 This shows a negligible contribution from the 2D Brillouin zone with a transmission coefficient of $10^{-10}$.

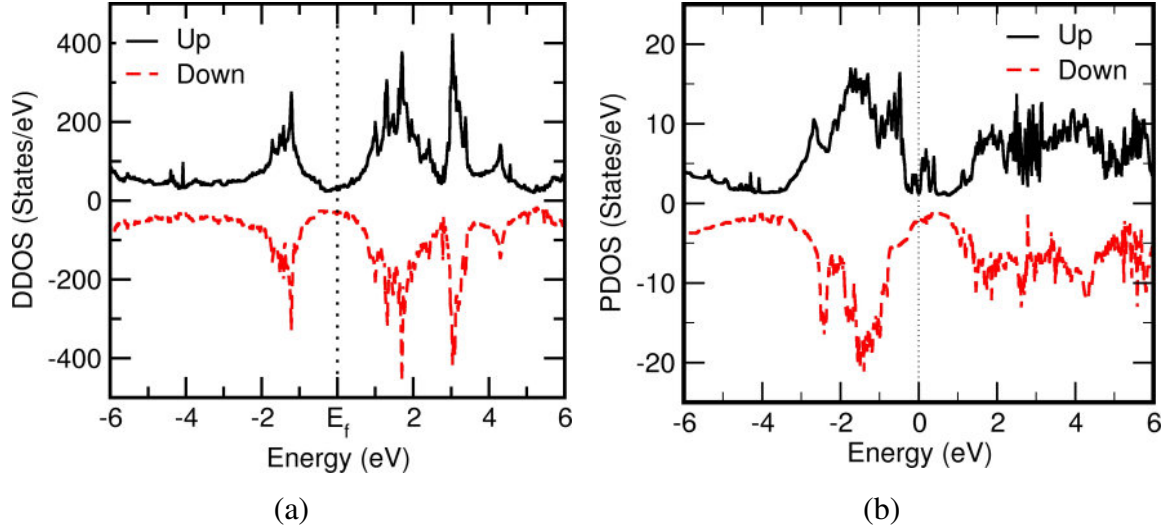

*Figure 25 Total device density of states, projected density of states of MoS$_2$ monolayer in anti-parallel orientation in NiMnSb/MoS$_2$/NiMnSb junction (heterostructure-2).*

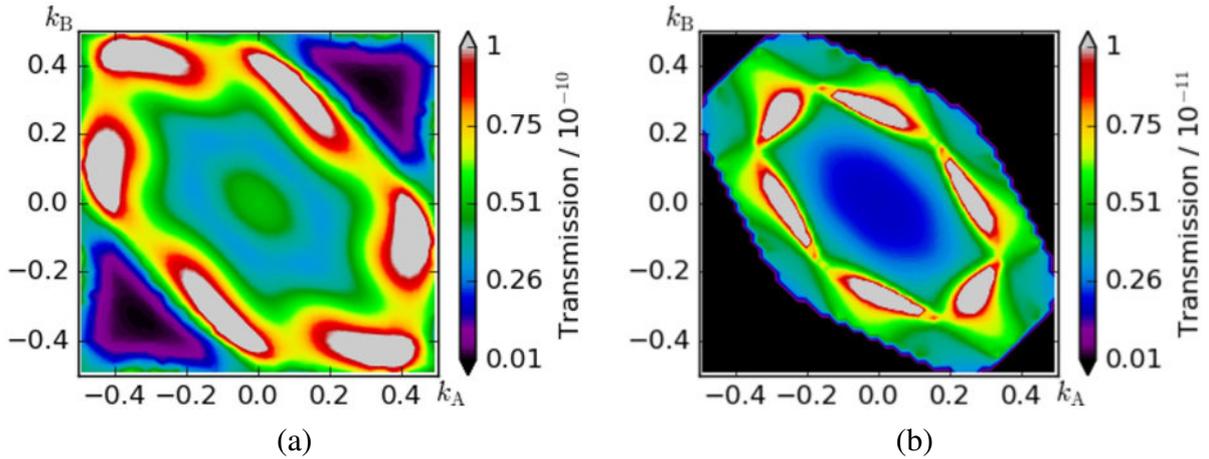

*Figure 26 K-resolved transmission spectrum of NiMnSb/MoS$_2$/NiMnSb device in anti-parallel configuration.*

The heterostructure-2 with single layer MoS$_2$ as spacer shows metallic behaviour; the spin-up channel is metallic due to the strong hybridisation with transition metals and the strong covalent bonding between the Mn and S atoms at the interface. Due to the metallic nature of the interface, finite transmission for the junction at fermi-level in zero-bias condition is seen. The finite bias calculation for the junction with a small bias of 0.05V for GMR measurement is carried out. The transmission spectrum for bias 0.05V in parallel configuration is shown in Figure 27(a), and the corresponding K-resolved transmission spectrum is shown in Figure 27 (c). The transmission spectrum for parallel configuration at bias 0.05V shows that the spin-up channel has a transmission of 0.7 at the fermi level and zero transmission in the spin-down channel. The transmission in the spin-up channel in the heterostructure-2 with a single layer of $MoS_2$ as spacer is higher than the transmission for heterostructure-1 with a single layer of $MoS_2$ as spacer. The K-resolved transmission spectrum in Figure 27 (c) shows K-points away from the Brillouin zone centre, forming a hexagonal ring structure that contributes to the fermi-level transmission. For anti-parallel configuration, the transmission is zero at the fermi-level, and there is no contribution from the Brillouin zone for K-resolved transmission, as seen in Figure 27 (b) and Figure 27 (d), respectively.

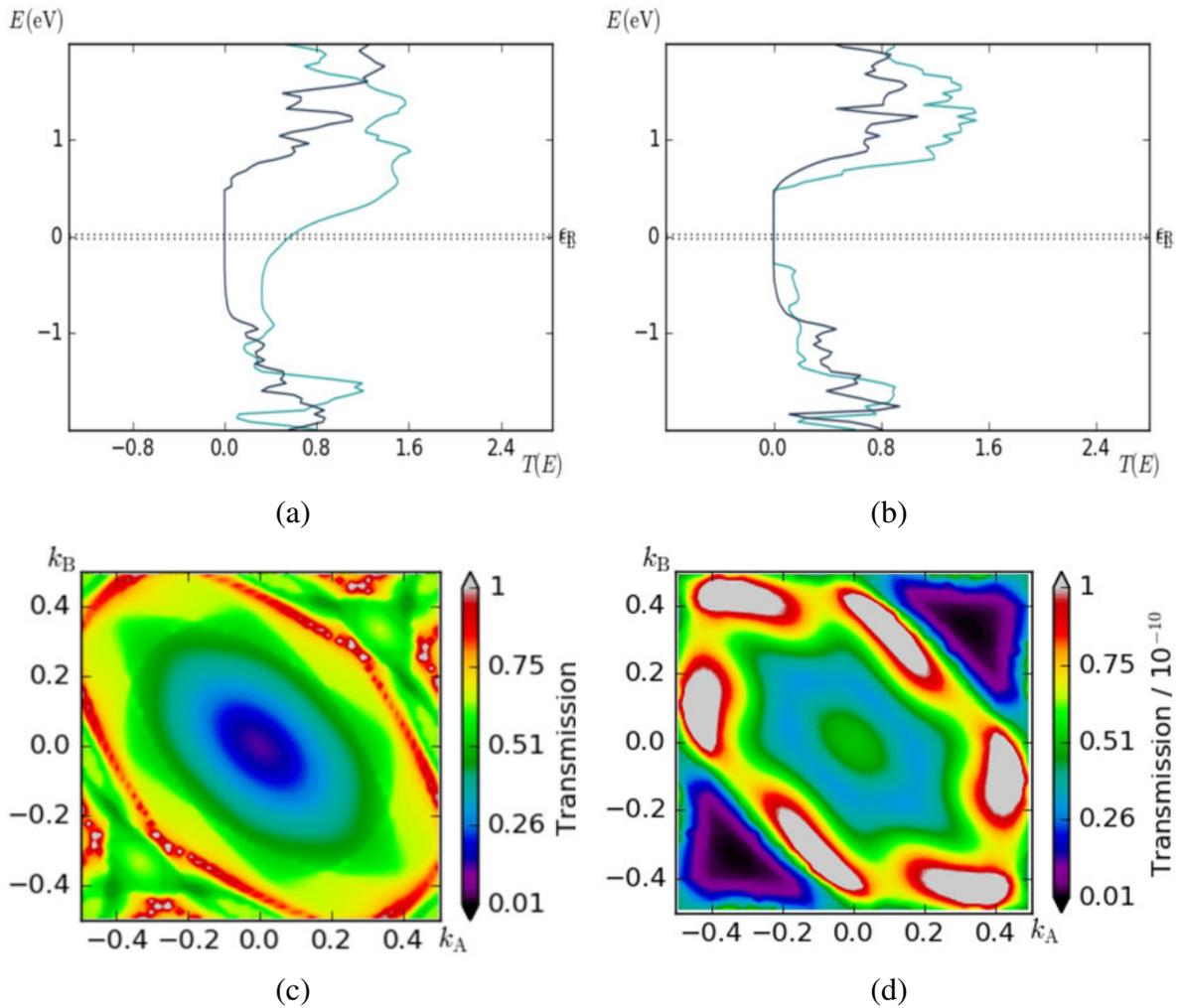

(a)      (b)

(c)      (d)

*Figure 27   Transmission spectrum and K-resolved transmission spectrum for bias 0.05V in parallel (left panel) and anti-parallel (right panel)*

*Three-layer MoS$_2$ spacer in heterostructure-2*

The NiMnSb/MoS$_2$/NiMnSb junction with three layers of MoS$_2$ as spacer in heterostructure-2. The interface is made by the Mn truncated NiMnSb electrode on one side of MoS$_2$ and the Sb truncated electrode on another side of the NiMnSb/MoS$_2$ junction, thereby forming Mn-S and Sb-S bonding at the interface. Figure 28 (a) the PDOS for the three-layer and middle layer MoS$_2$ with Mn-S and Sb-S bonding at the junction. The PDOS for the three layers indicates that the DOS has few states in the spin-up channel at the fermi level of the junction, making it metallic. The heterostructure-2 with Mn-S and Sb-S bond has more states at the fermi level than the interface made from Sb-S and Sb-S bonding at the junction. However, The PDOS for the middle layer of three-layer MoS$_2$ remains semiconducting/has no states at the Fermi level.

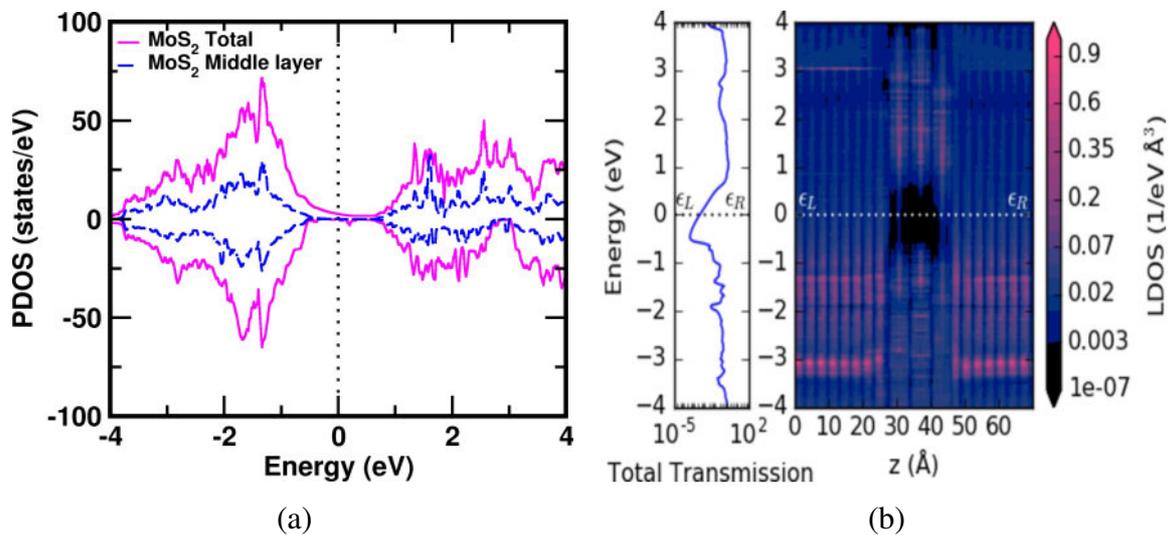

*Figure 28   Projected density of states of MoS$_2$ 3-layer and middle layer MoS$_2$ figure (a), projected device local density of states figure (b).*

The PLDOS for the device with heterostructure-2 is shown in Figure 29 (b). The metallicity of the Mn-S bonded interface is also seen from the PLDOS, where the availability of states at the Mn-S bonded is seen. This shows the strong covalent bonding of the Mn-S bond. At the Sb-S bonded side of the interface, the Sb-S bonding is weaker than the Mn-S bonding and is also seen from the PLDOS at the Sb-S bonded side of the interface. Where the LDOS touches the fermi level in the spin-up channel. However, the middle layer of the three-layer MoS$_2$ does not contribute to the DOS, as seen from the PLDOS and PDOS.

The spin-resolved electron transmission for parallel and anti-parallel configuration at zero bias is shown in Figure 29 (a) and Figure 29 (b) respectively. For parallel configuration, the transmission for the spin-up and spin-down channels is zero at the fermi level. For the anti-parallel configuration, the transmission spectrum for both channels is identical and has no transmission of electrons at the zero bias.

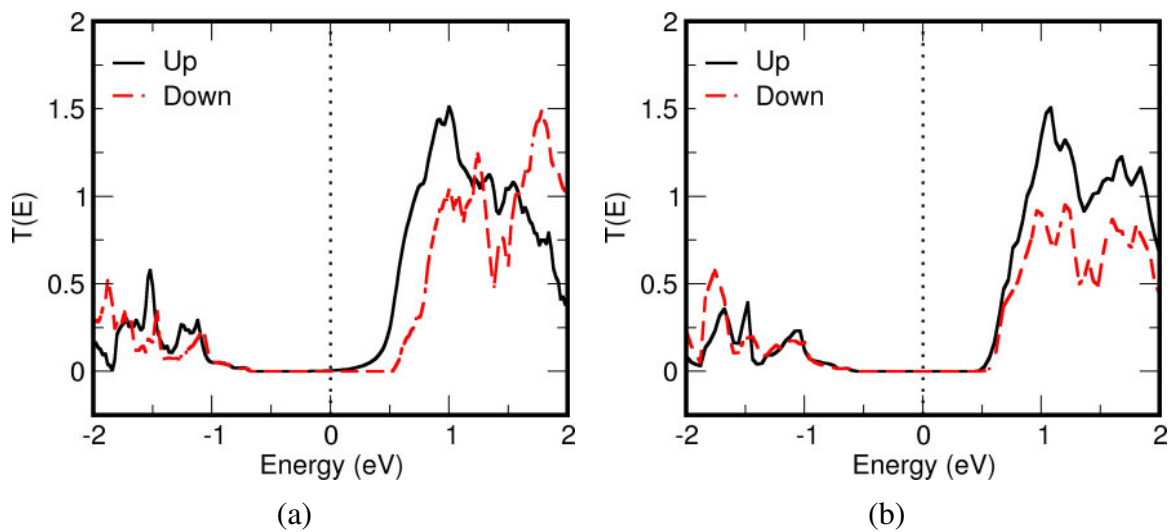

(a)  (b)

*Figure 29  Transmission spectrum along spin-up and spin-down channel of NiMnSb/MoS$_2$(3-layer)/NiMnSb (heterostructure-2) in parallel and anti-parallel orientation of electrodes.*

Figure 30 shows the K-resolved spin-depended transmission spectrum for parallel and anti-parallel device configurations. The spin-dependent electron transmission for the spin-up channel in Figure 30 (a) shows six packets from the Brillouin zone are contributing to the transmission of electrons through the channel, with maximum transmission co-efficient of the order of 10$^{-3}$ from the six pockets in hexagonal 2D Brillouin zone. The K-resolved transmission spectrum for the spin-down channel in Figure 30 (b) shows no contribution from the 2D Brillouin zone to the transmission as it has zero transmission, as seen in Figure 29 (a). The K-resolved transmission spectrum for the anti-parallel configuration of the device in Figure 30 (c) and 30 (d) shows that both spin-up and spin-down channels do not contribute to the transmission at the fermi-level. For anti-parallel device configuration, the transmission coefficient is of the order of $10^{-10}$ from the 2D Brillouin zone.

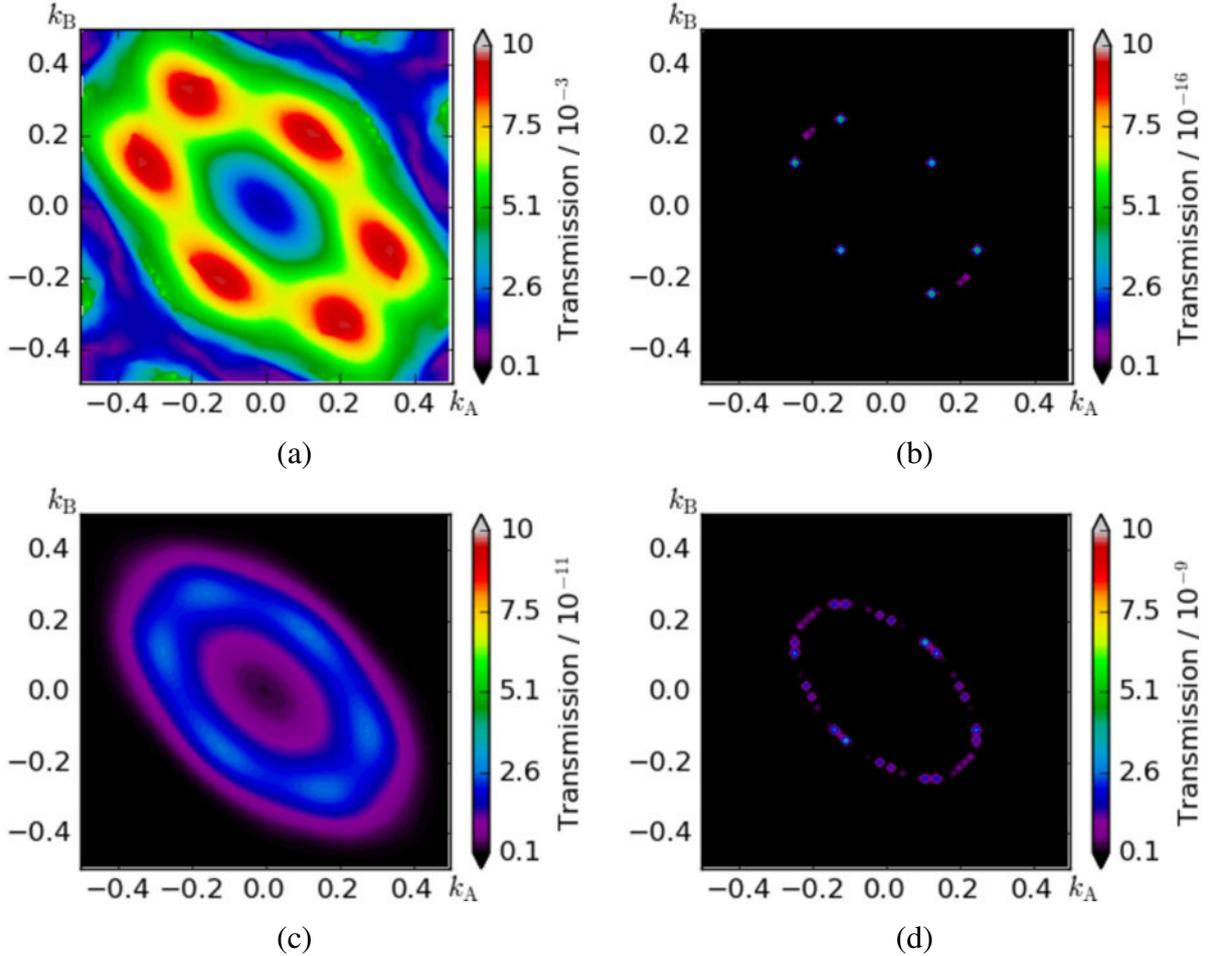

*Figure 30    K-resolved transmission spectrum for the device in parallel and anti-parallel configuration in NiMnSb/MoS$_2$(3-layer)/NiMnSb (heterostructure-2).*

## Measurement of Conductance

Conductance measurements for the parallel and anti-parallel device configurations for the different heterostructures for the measurement of GMR and TMR ratios are calculated at zero bias, and a slight bias of 0.05V is presented in Table 1. The conductance measurement at zero bias for the heterostructures shows a GMR of 100% for the devices, as the conductance is only due to one spin channel.

*Table 1    Conductance measurement for different heterostructures with monolayer and 3-layer MoS₂ as spacer.*

| Heterostructures | | | Monolayer Zero bias | Monolayer Bias 0.05V | 3layer Zero bias | 3layer Bias 0.05V |
|---|---|---|---|---|---|---|
| Heterostructure-1 Conductance (Siemens) | P | UP | 1.83e-06 | 1.95e-06 | 9.56e-09 | 9.66e-09 |
| | | Down | 6.98e-23 | 8.74e-20 | 7.43e-27 | 5.03e-27 |
| | | Total | 1.83e-06 | 1.95e-06 | 9.56e-09 | 9.66e-09 |
| | AP | UP | 1.88e-13 | 4.35e-14 | 1.20e-18 | 1.98e-12 |
| | | Down | 9.15e-16 | 3.46e-13 | 9.01e-16 | 7.92e-19 |
| | | Total | 1.89e-13 | 3.89e-13 | 9.02e-16 | 1.98e-12 |
| | GMR | | 100.00 % | 100% | 100.00 % | 99.96 % |
| Heterostructure-2 Conductance (Siemens) | P | Up | 2.52e-05 | 2.22e-05 | 1.94e-07 | 9.96e-08 |
| | | Down | 3.01e-23 | 8.39e-25 | 3.54e-15 | 2.37e-22 |
| | | Total | 2.52e-05 | 2.22e-05 | 1.94e-07 | 9.96e-08 |
| | AP | Up | 2.10e-15 | 6.85e-11 | 2.71e-16 | 3.73e-09 |
| | | Down | 1.21e-16 | 3.57e-13 | 5.84e-16 | 7.68e-19 |
| | | Total | 2.23e-15 | 6.88e-11 | 8.55e-16 | 3.73e-09 |
| | GMR | | 100.00 % | 100% | 100.00 % | 92.78 % |

Conclusions

Spin-dependent electron transport in MTJ using half-metallic Heusler alloy NiMnSb as electrode for efficient spin injection in MoS$_2$ as spacer is studied. We have studied the electron transport along the (0001) direction of MoS$_2$ with two different thickness; Single layer MoS$_2$ (SL) and three-layer MoS$_2$ (3L) and two different types (Mn-S and Sb-S atoms making bonds at the interface) of interface geometries based on Sb and Mn truncated NiMnSb(111) surfaces as electrodes and MoS$_2$(0001) surfaces as spacers. The heterostructures with monolayer MoS$_2$ as spacer are metallic in the spin-up channel; the interface is metallic due to the strong hybridisation of *d*-orbitals of transition metal Ni, Mn and Mo atoms. The half-metallic NiMnSb keeps its half-metallic nature at the interface in the heterostructures. The interface effect is limited to two layers of NiMnSb and an adjacent layer of $MoS_2$ only at the interface. The junction with monolayer $MoS_2$ is nearly half-metallic; the spin-up channel is metallic, and the spin-down channel has a narrow band gap. In the junction with three layers of $MoS_2$, the middle layer of $MoS_2$ keeps its semiconducting nature; this is the effect of the interface limited to the first layer of $MoS_2$ only. The junction with interface formed by Mn truncated NiMnSb electrode, and $MoS_2$ has stronger hybridisation due to strong covalent bonding between Mn and S, leading to a more conducting nature of the device. The junction with the interface formed by the Sb truncated NiMnSb electrode and $MoS_2$ has weaker bonding between the Sb and S atoms; thus, the junction conducts less in the spin-up channel. The transmission spectrum for the devices shows transmission only due to the spin-up channel. Hence, injection of 100% spin-polarised current is obtained. The device can act as a spin filter or spin switch where the current conduction is only due to one type of charge carrier.

# References

[1]  M. Julliere, Tunneling between ferromagnetic films, Phys Lett A **54**, 225 (1975).

[2]  M. N. Baibich, J. M. Broto, A. Fert, F. N. Van Dau, F. Petroff, P. Eitenne, G. Creuzet, A. Friederich, and J. Chazelas, Giant magnetoresistance of (001)Fe/(001)Cr magnetic superlattices, Phys Rev Lett **61**, 2472 (1988).

[3]  G. Binasch, P. Grünberg, F. Saurenbach, and W. Zinn, Enhanced magnetoresistance in layered magnetic structures with antiferromagnetic interlayer exchange, Phys Rev B **39**, 4828 (1989).

[4]  A. Hirohata and K. Takanashi, Future perspectives for spintronic devices, J Phys D Appl Phys **47**, (2014).

[5]  J. P. Degrave, A. L. Schmitt, R. S. Selinsky, J. M. Higgins, D. J. Keavney, and S. Jin, Spin polarization measurement of homogeneously doped $Fe_{1-x}Co_xSi$ nanowires by andreev reflection spectroscopy, Nano Lett **11**, 4431 (2011).

[6]  L. Zhang, J. Zhou, H. Li, L. Shen, and Y. P. Feng, Recent Progress and Challenges in Magnetic Tunnel Junctions with 2D Materials for Spintronic Applications, Vol. 8 (2021).

[7]  S. Bhatti, R. Sbiaa, A. Hirohata, H. Ohno, S. Fukami, and S. N. Piramanayagam, Spintronics based random access memory: a review, Materials Today **20**, 530 (2017).

[8]  de G. R. A, M. F. M, van E. P. G, and B. K. H. J, New class of materials: half-metallic ferromagnets, Phys. Rev. Lett. **50**, 2024 (1983).

[9]  I. Galanakis and P. Mavropoulos, Spin-polarization and electronic properties of half-metallic Heusler alloys calculated from first principles, Journal of Physics: Condensed Matter **19**, 315213 (2007).

[10]  I. Galanakis, Surface properties of the half-and full-Heusler alloys, Journal of Physics: Condensed Matter **14**, 6329 (2002).

[11]  W. E. Pickett and D. J. Singh, Electronic structure and half-metallic transport in the La, Phys Rev B **53**, 1146 (1996).

[12]  S. J. Youn and B. I. Min, Effects of the spin-orbit interaction in Heusler compounds: Electronic structures and Fermi surfaces of NiMnSb and PtMnSb, Phys Rev B **51**, 10436 (1995).

[13]  A. Yanase and N. Hamada, Electronic Structure in High Temperature Phase of $Fe_3O_4$, JPSJ.68.1607 **68**, 1607 (2013).

[14]  K. Schwarz, $CrO_2$ predicted as a half-metallic ferromagnet, Journal of Physics F: Metal Physics **16**, L211 (1986).

[15]  X. Li, X. Wu, and J. Yang, Half-metallicity in $MnPSe_3$ exfoliated nanosheet with carrier doping, J Am Chem Soc **136**, 11065 (2014).

[16]  E. Torun, H. Sahin, S. K. Singh, and F. M. Peeters, Stable half-metallic monolayers of $FeCl_2$, Appl Phys Lett **106**, (2015).

[17] J. Y. Chen, X. X. Li, W. Z. Zhou, J. L. Yang, F. P. Ouyang, and X. Xiong, Large-Spin-Gap Nodal-Line Half-Metal and High-Temperature Ferromagnetic Semiconductor in $Cr_2X_3$ (X=O,S,Se) Monolayers, Adv Electron Mater **6**, 1900490 (2020).

[18] X. Zhang, B. Wang, Y. Guo, Y. Zhang, Y. Chen, and J. Wang, High Curie temperature and intrinsic ferromagnetic half-metallicity in two-dimensional $Cr_3X_4$ (X = S, Se, Te) nanosheets, Nanoscale Horiz **4**, 859 (2019).

[19] Y. Shen, D. Kan, I. C. Lin, M. W. Chu, I. Suzuki, and Y. Shimakawa, Perpendicular magnetic tunnel junctions based on half-metallic $NiCo_2O_4$, Appl Phys Lett **117**, (2020).

[20] M. J. Otto, H. Feil, R. A. M. Van Woerden, J. Wijngaard, P. J. Van Der Valk, C. F. Van Bruggen, and C. Haas, Electronic structure and magnetic, electrical and optical properties of ferromagnetic Heusler alloys, J Magn Magn Mater **70**, 33 (1987).

[21] R. B. Helmholdt, R. A. de Groot, F. M. Mueller, P. G. van Engen, and K. H. J. Buschow, Magnetic and crystallographic properties of several C1b type Heusler compounds, J Magn Magn Mater **43**, 249 (1984).

[22] M. J. Otto, R. A. M. Van Woerden, P. J. Van Der Valk, J. Wijngaard, C. F. Van Bruggen, C. Haas, and K. H. J. Buschow, Half-metallic ferromagnets. I. Structure and magnetic properties of NiMnSb and related inter-metallic compounds, Journal of Physics: Condensed Matter **1**, 2341 (1989).

[23] J. Y. Chen, Y. C. Lau, J. M. D. Coey, M. Li, and J. P. Wang, High Performance MgO-barrier Magnetic Tunnel Junctions for Flexible and Wearable Spintronic Applications, Scientific Reports **7**, 1 (2017).

[24] B. Canto, C. P. Gouvea, B. S. Archanjo, J. E. Schmidt, and D. L. Baptista, On the Structural and Chemical Characteristics of $Co/Al_2O_3$/graphene Interfaces for Graphene Spintronic Devices, Scientific Reports **5**, 1 (2015).

[25] B. Taudul et al., Tunneling Spintronics across MgO Driven by Double Oxygen Vacancies, Adv Electron Mater **3**, 1600390 (2017).

[26] Ikhtiar, S. Kasai, P. H. Cheng, T. Ohkubo, Y. K. Takahashi, T. Furubayashi, and K. Hono, Magnetic tunnel junctions with a rock-salt-type $Mg_{1-x}Ti_xO$ barrier for low resistance area product, Appl Phys Lett **108**, (2016).

[27] M. Schäfers, V. Drewello, G. Reiss, A. Thomas, K. Thiel, G. Eilers, M. Münzenberg, H. Schuhmann, and M. Seibt, Electric breakdown in ultrathin MgO tunnel barrier junctions for spin-transfer torque switching, Appl Phys Lett **95**, (2009).

[28] H. X. Wei, Q. H. Qin, M. Ma, R. Sharif, and X. F. Han, 80% tunneling magnetoresistance at room temperature for thin Al-O barrier magnetic tunnel junction with CoFeB as free and reference layers, J Appl Phys **101**, (2007).

[29] N. Tombros, C. Jozsa, M. Popinciuc, H. T. Jonkman, and B. J. Van Wees, Electronic spin transport and spin precession in single graphene layers at room temperature, Nature **448**, 571 (2007).

[30] M. V. Kamalakar, C. Groenveld, A. Dankert, and S. P. Dash, Long distance spin communication in chemical vapour deposited graphene, Nature Communications **6**, 1 (2015).

[31] M. V. Kamalakar, A. Dankert, P. J. Kelly, and S. P. Dash, Inversion of Spin Signal and Spin Filtering in Ferromagnet|Hexagonal Boron Nitride-Graphene van der Waals Heterostructures, Scientific Reports **6**, 1 (2016).

[32] Z. Gong, G. Bin Liu, H. Yu, D. Xiao, X. Cui, X. Xu, and W. Yao, Magnetoelectric effects and valley-controlled spin quantum gates in transition metal dichalcogenide bilayers, Nature Communications **4**, 1 (2013).

[33] C. H. Li, O. M. J. Van't Erve, J. T. Robinson, Y. Liu, L. Li, and B. T. Jonker, Electrical detection of charge-current-induced spin polarization due to spin-momentum locking in $Bi_2Se_3$, Nat Nanotechnol **9**, 218 (2014).

[34] W. Li, L. Xue, H. D. Abruña, and D. C. Ralph, Magnetic tunnel junctions with single-layer-graphene tunnel barriers, Phys Rev B Condens Matter Mater Phys **89**, 184418 (2014).

[35] O. V. Yazyev and A. Pasquarello, Magnetoresistive junctions based on epitaxial graphene and hexagonal boron nitride, Phys Rev B Condens Matter Mater Phys **80**, 035408 (2009).

[36] H. Zhou, Y. Zhang, and W. Zhao, Tunable Tunneling Magnetoresistance in van der Waals Magnetic Tunnel Junctions with 1 T -$CrTe_2$ Electrodes, ACS Appl Mater Interfaces **13**, 1214 (2021).

[37] A. Splendiani, L. Sun, Y. Zhang, T. Li, J. Kim, C. Y. Chim, G. Galli, and F. Wang, Emerging photoluminescence in monolayer $MoS_2$, Nano Lett **10**, 1271 (2010).

[38] J. K. Ellis, M. J. Lucero, and G. E. Scuseria, The indirect to direct band gap transition in multilayered $MoS_2$ as predicted by screened hybrid density functional theory, Appl Phys Lett **99**, (2011).

[39] K. Dolui, A. Narayan, I. Rungger, and S. Sanvito, Efficient spin injection and giant magnetoresistance in Fe/$MoS_2$/Fe junctions, Phys Rev B Condens Matter Mater Phys **90**, 1 (2014).

[40] H. C. Wu et al., Spin-dependent transport properties of $Fe_3O_4$/$MoS_2$/$Fe_3O_4$ junctions, Sci Rep **5**, 1 (2015).

[41] W. Wang et al., Spin-Valve Effect in NiFe/$MoS_2$/NiFe Junctions, Nano Lett **15**, 5261 (2015).

[42] L. F. Mattheiss, Band structures of transition-metal-dichalcogenide layer compounds, Phys Rev B **8**, 3719 (1973).

[43] R. Coehoorn, C. Haas, J. Dijkstra, C. J. F. Flipse, R. A. De Groot, and A. Wold, Electronic structure of $MoSe_2$, $MoS_2$, and $WSe_2$. I. Band-structure calculations and photoelectron spectroscopy, Phys Rev B **35**, 6195 (1987).

[44] B. Radisavljevic, A. Radenovic, J. Brivio, V. Giacometti, and A. Kis, Single-layer $MoS_2$ transistors, Nat Nanotechnol **6**, 147 (2011).

[45] S. Lebègue and O. Eriksson, Electronic structure of two-dimensional crystals from ab initio theory, Phys Rev B Condens Matter Mater Phys **79**, 4 (2009).

[46] K. F. Mak, C. Lee, J. Hone, J. Shan, and T. F. Heinz, Atomically thin $MoS_2$: A new direct-gap semiconductor, Phys Rev Lett **105**, 2 (2010).

[47] R. B. Murray and A. D. Yoffe, The band structures of some transition metal dichalcogenides: Band structures of the titanium dichalcogenides, Journal of Physics C: Solid State Physics **5**, 3038 (1972).

[48] L. Wei, C. Jun-fang, H. Qinyu, and W. Teng, Electronic and elastic properties of $MoS_2$, Physica B Condens Matter **405**, 2498 (2010).

[49] Z. Y. Zhu, Y. C. Cheng, and U. Schwingenschlögl, Giant spin-orbit-induced spin splitting in two-dimensional transition-metal dichalcogenide semiconductors, Phys Rev B Condens Matter Mater Phys **84**, 153402 (2011).

[50] K. Endo, Magnetic Studies of Clb-Compounds CuMnSb, PdMnSb and $Cu_{1-x}$(Ni or Pd)$_x$ MnSb, JPSJ **29**, 643 (2013).

[51] K. Tarawneh, N. Al-Aqtash, and R. Sabirianov, Large magnetoresistance in planar Fe/$MoS_2$/Fe tunnel junction, Comput Mater Sci **124**, 15 (2016).

[52] W. Rotjanapittayakul, W. Pijitrojana, T. Archer, S. Sanvito, and J. Prasongkit, Spin injection and magnetoresistance in $MoS_2$-based tunnel junctions using $Fe_3Si$ Heusler alloy electrodes, Scientific Reports **8**, 1 (2018).

[53] R. A. de Groot and G. A. de Wijs, Towards 100% spin-polarized charge-injection: The half-metallic NiMnSb/CdS interface, Phys Rev B **64**, 020402 (2001).

[54] S. J. Jenkins, Ternary half-metallics and related binary compounds: Stoichiometry, surface states, and spin, Phys Rev B Condens Matter Mater Phys **70**, 1 (2004).

[55] M. Ležaić, P. Mavropoulos, G. Bihlmayer, and S. Blügel, Scanning tunnelling microscopy of surfaces of half-metals: an ab−initio study on NiMnSb(001), J Phys D Appl Phys **39**, 797 (2006).

[56] A. Debernardi, M. Peressi, and A. Baldereschi, Spin polarization and band alignments at NiMnSb/GaAs interface, Comput Mater Sci **33**, 263 (2005).

[57] I. Galanakis, M. Ležaić, G. Bihlmayer, and S. Blügel, Interface properties of NiMnSb InP and NiMnSb GaAs contacts, Phys Rev B Condens Matter Mater Phys **71**, 214431 (2005).

[58] S. Smidstrup et al., QuantumATK: An integrated platform of electronic and atomic-scale modelling tools, J. Phys: Condens. Matter **32**, 15901 (2020).

[59] M. Brandbyge, J.-L. Mozos, P. Ordejón, J. Taylor, and K. Stokbro, Density-functional method for nonequilibrium electron transport, Phys Rev B **65**, 165401 (2002).

[60] D. Stradi, U. Martinez, A. Blom, M. Brandbyge, and K. Stokbro, General atomistic approach for modeling metal-semiconductor interfaces using density functional theory and nonequilibrium Green's function, Phys Rev B **93**, 155302 (2016).

[61] S. Smidstrup, D. Stradi, J. Wellendorff, P. A. Khomyakov, U. G. Vej-Hansen, M.-E. Lee, T. Ghosh, E. Jónsson, H. Jónsson, and K. Stokbro, First-principles Green's-function method for surface calculations: A pseudopotential localized basis set approach, Phys Rev B **96**, 195309 (2017).

[62] J. M. Soler, E. Artacho, J. D. Gale, A. García, J. Junquera, P. Ordejón, and D. Sánchez-Portal, The SIESTA method for ab initio order-N materials simulation, Journal of Physics: Condensed Matter **14**, 2745 (2002).

[63] J. P. Perdew, K. Burke, and M. Ernzerhof, Generalized Gradient Approximation Made Simple, Phys Rev Lett **77**, 3865 (1996).

[64] H. J. Monkhorst and J. D. Pack, Special points for Brillouin-zone integrations, Phys Rev B **13**, 5188 (1976).

[65] D. Stradi, L. Jelver, S. Smidstrup, and K. Stokbro, Method for determining optimal supercell representation of interfaces, Journal of Physics: Condensed Matter **29**, 185901 (2017).

[66] S. Grimme, J. Antony, S. Ehrlich, and H. Krieg, A consistent and accurate ab initio parametrization of density functional dispersion correction (DFT-D) for the 94 elements H-Pu, J Chem Phys **132**, 154104 (2010).